\PassOptionsToPackage{sort&compress,numbers,super}{natbib}
\documentclass[a4paper,amsfonts,amssymb,amsmath,reprint,showkeys,toside,hidelinks]{revtex4-1}

\usepackage[english]{babel}
\usepackage{amsthm}
\usepackage{mathtools}
\usepackage{physics}
\usepackage{xcolor}
\usepackage{graphics}
\usepackage{graphicx}
\usepackage{adjustbox}
\usepackage{placeins}
\usepackage[T1]{fontenc}
\usepackage{lipsum}
\usepackage{csquotes}
\usepackage{tabularx}
\usepackage{ragged2e}
\usepackage{makecell}
\setcitestyle{super}
\usepackage{booktabs}
\usepackage{siunitx}
\usepackage[etalmode=truncate,maxauthors=0,articletitle=true,doi=true]{achemso}

\AtBeginDocument{}

\usepackage[left=23mm,right=13mm,top=35mm,columnsep=15pt]{geometry}
\usepackage[pdftex,pdftitle={Article},pdfauthor={Author}]{hyperref}

\graphicspath{Images}

\mciteErrorOnUnknownfalse

\makeatletter
\AtBeginDocument{%
  \let\SM@orig@label\label
  \protected\def\label#1{%
    \SM@orig@label{#1}%
    \SM@orig@label{S-#1}%
    \SM@orig@label{M-#1}%
  }%
}
\makeatother

\begin{document}

\title{Self-Assembled Naphthalocyanine Molecular Lattices as a Platform for Tunable Quantum Superlattices}

\author{Adrian Bahri$^1$}
\thanks{These authors contributed equally to this work.}
\author{Zhibo Kang$^2$}
\thanks{These authors contributed equally to this work.}
\author{Ziyan Zhu$^3$}
\author{Eric I. Altman$^{4,5}$}
\author{Yu He$^2$}
\email{yu.he@yale.edu}
\author{Chunjing Jia$^1$}
\email{chunjing@phys.ufl.edu}

\affiliation{$^1$Department of Physics, University of Florida, 2001 Museum Rd, Gainesville, FL 32611\\
    $^2$ Department of Applied Physics, Yale University, New Haven, CT 06511\\
    $^3$ Department of Physics, Boston College, 
Chestnut Hill, MA 02467\\    
    $^4$ Department of Chemical and Environmental Engineering, Yale University, New Haven, CT 06511\\    
    $^5$ Department of Materials Science, Yale University, New Haven, CT 06511}

\begin{abstract}
\noindent Compared to van der Waals moiré systems, molecular assembly has emerged as an exciting alternative platform for superlattice engineering via heterointegration. The electronic properties of the self-assembled square lattice monolayer molecular crystal of metal-free naphthalocyanine (H$_2$Nc), in particular the electronic band dispersion and their tunability by metal substrates, remain less explored. Using density functional theory, supported by angle-resolved photoemission and scanning tunneling microscopy, we compare the electronic structure of a free-standing H$_2$Nc monolayer with that of H$_2$Nc lattice assembled on noble metal substrates. In the free-standing film, we identify both nearly flat, molecule-localized states and more dispersive bands, and we show that each can be compactly described by an anisotropic tight-binding Hamiltonian that yields band-resolved hopping anisotropies. We further show theoretically the wide tunability in the inter-site hopping and Coulomb interaction based on dielectric and screening environment, and show two experimental realizations using Ag(100) and Au(111) substrates. On Ag(100), orbital hybridization between molecular frontier states and the substrate drives finite spectral weight at the Fermi level and local interfacial polarization. This hybridization enhances the effective intermolecular hopping roughly thirty-fold, drastically reducing $U/t$. Complementary angle-resolved photoemission spectroscopy resolves substantial interfacial charge redistribution and enhanced bandwidth of the HOMO band, consistent with theoretical predictions. Although static ground-state calculations exhibit an intrinsic $C_2$ hopping anisotropy set by the two central hydrogens, thermal averaging over tautomer configurations at liquid nitrogen temperature (77~K) restores an effectively $C_4$-symmetric dispersion while preserving the overall bandwidth. These results clarify how metal substrates and gates convert H$_2$Nc from isolated molecules into a tunable 2D lattice, and highlight molecular superlattices as a promising platform in which the effective interaction parameters can be engineered over a wide range.

\end{abstract}


\maketitle

%=====================================================================
%  MAIN TEXT
%=====================================================================

%\section*{Introduction}
\label{sec:intro}

Moiré engineering enables a new route to quantum simulation of correlated electrons by using stacked atomic layers to create artificial superlattices. It successfully enables the simulation of topological and correlated physics such as the integer and fractional quantum anomalous Hall effect~\cite{IQAHEandFQAHEinTBMoTe22023,FQAHE2023,FQAHEinMoireMoTe22023}.  However, the overall low energy scale, spatially dependent interlayer spacing,~\cite{Magic2021} structural domains,~\cite{TBLG2019} and twist angle disorder~\cite{uri2020mapping} have made investigations at the microscopic Hamiltonian level challenging. Recent advances such as robotic stacking and hBN encapsulation \cite{moireRoboticStacking,moirehBNencap}, however, have improved twist-angle uniformity, enabling a plethora of quantum simulation studies.

Instead of the traditional top-down approach of mechanically stacking and twisting atomic layers, molecular frameworks (MOFs, COFs, HOFs) and supramolecular self-assembly (SSA) on solid-state surfaces offer an alternative bottom-up route featuring artificial molecular superlattices with intrinsic structural order~\cite{ReviewArtificialLattices2022,ArtificialKagomeLattice2024,ArtificialLiebLattice2024,EngineeringQuantumStates2022}. This approach potentially enables functional systems with nanometer dimensions,
where traditional fabrication methods of the microelectronics industry reach their fundamental limits~\cite{MolecularNanostructures}. Molecular superlattices have also been reported to self-assemble with well-defined long-range periodicity,~\cite{SSrenormalization2019,IPZ2019c,Ules2014,Zhou2020,chowdhury2025bright} offering an inherently correlated platform with large interaction energy scales. However, most bulk molecular crystals are based on van der Waals coupling, which leads to very small inter-site hopping. In most cases, the molecular superlattice weakly perturbs the itinerant substrate electronic states, causing benign hybridization gaps. On the other hand, the molecular network itself can act as the artificial atomic lattice (``quantum dot array''), while the substrate provides means to control the intermolecular electron hopping.~\cite{SSrenormalization2019,EngineeringQuantumStates2022} This offers a complementary route to simulate lattice models from the local limit, and is of practical relevance for molecular network-based surface interconnects and conducting films without needing manual molecule-by-molecule placement.~\cite{Shang2015,Slot2017,ArtificialLiebLattice2024,ArtificialKagomeLattice2024} Here we address a matter-of-principle question: although molecular superlattices often weakly perturb itinerant substrate states, the degree of tunability of the electronic band dispersion within the two-dimensional molecular lattice itself remains unclear, as does the extent to which substrate electronic states renormalize its properties. In particular, we quantify the energy scale and anisotropy of intermolecular hopping in molecular superlattices and determine how these parameters can be tuned through coupling to metal substrates.

Among the most promising building blocks for such molecular lattices are conjugated macrocycles such as phthalocyanines and naphthalocyanines, which combine strong optical activity with chemically tunable frontier orbitals, making them attractive for use in nanoscale electronics, photonics, and sensing.~\cite{PcsReview1,PcsReview2,OrganicElectronicsReview} In real devices, however, the electronic character of an isolated molecule is often dramatically reshaped by metal contacts: adsorption can change level alignment, induce charge transfer and screening, and produce new substrate-mediated electronic states.~\cite{PcsSTM,PcsARPES,PcsOnSubstrate} Because contact-driven modifications can qualitatively alter transport and optical response, a quantitative understanding of how metal substrates change the electronic properties of such molecular monolayers is essential for device design and for interpreting spectroscopic measurements. Experimental and theoretical studies of phthalocyanines and related macrocycles on noble metals have documented a range of behaviors from weak physisorption to stronger hybridization and partial charge transfer at the single-molecule level.~\cite{Li2010,Kroger2011,Granet2020,Wu2019,Bai2008} However, collective film properties, such as band formation, dispersion, and partial filling, cannot be inferred from single-molecule probes alone. 

In this work, we report a combined theoretical and experimental investigation on the electronic structure of H$_2$Nc molecular superlattices on noble metal substrates. Compared to its metallated siblings, H$_2$Nc is a simpler system without the complication of the additional $d$ states from the center metal atoms. Moreover, in contrast to the smaller H$_2$Pc, H$_2$Nc has a more stable assembly at room temperature and is more resilient to higher temperature treatment.~\cite{Naph}
Using first principles methods, we investigate the ground state electronic symmetry, hopping anisotropy, and on-site Coulomb repulsions in the free-standing monolayer. We then study the charge transfer and band hybridization effects when the molecules assemble on a noble metal substrate. Finally, we focus on angle-resolved photoemission spectroscopy (ARPES) with the aid of scanning tunneling microscopy (STM) to examine the theoretical predictions for the H$_2$Nc molecular superlattice on Ag(100) and Au(111), and reveal large charge transfer and HOMO bandwidth enhancements. Validated by both theory and experiment, the length and energy scales of such 2D molecular monolayers closely match those of moiré lattices. This alignment, combined with the ability to gate into diverse electronic bands and an expansive library of molecule-substrate combinations, establishes these networks as a highly tunable platform for quantum simulation that extends beyond current moiré systems.

\section*{Results}
\label{sec:results and discussion}

\subsection{Theoretical electronic structure of the H$_2$Nc monolayer}

\begin{figure*}
\includegraphics[width=.85\textwidth]{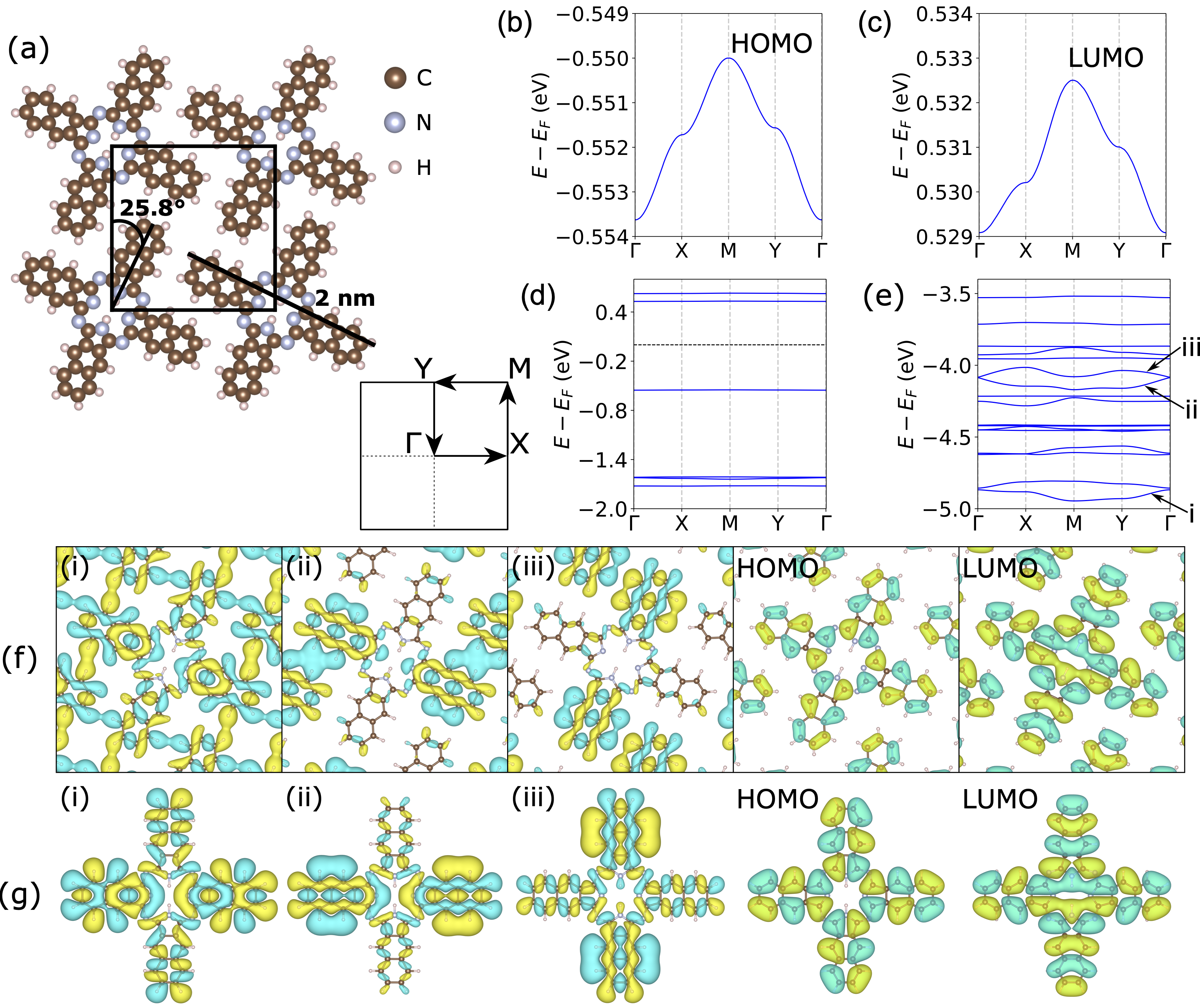}  
\caption{(a) Top view of the relaxed H$_2$Nc thin film and high-symmetry path through the first Brillouin zone of the anisotropic square lattice. (b,c) Close-up views of the (b) HOMO and (c) LUMO bands near the Fermi level. (d) Wide-window band structure illustrating the isolated nature of the HOMO band, bounded by the deeper occupied manifold below and the LUMO/LUMO+1 bands above. (e) Deeper energy spectrum; labels i--iii mark the most dispersive lower-energy bands. (f,g) Real parts of the $\Gamma$-point Bloch orbitals for (f) the thin film and (g) an isolated molecule. Orbitals were generated using \textsc{vaspkit}~\cite{vaspkit} and visualized using \textsc{VESTA}.~\cite{VESTA}}
\label{fig:bands_and_wfns}
\end{figure*} 

We begin with a theoretical investigation of the atomic and electronic structure of free-standing H$_2$Nc 2D lattices. We focus on the arrangement where all H$_2$Nc molecules are oriented along the same direction, so each unit cell contains a single H$_2$Nc molecule, which we found to be the most energetically favorable configuration. Alternative tautomer orderings, which result in different overall lattice symmetries, are presented in SI Section~\ref{S-sec:struct_char}. As seen in Fig.~\ref{fig:bands_and_wfns} (a), the H$_2$Nc molecule has two central hydrogen atoms that reduce the nominal $C_4$ symmetry of the macrocycle to $C_2$, resulting in a built-in anisotropy for many electronic orbitals in the thin-film. While the lattice is nearly square ($a_y/a_x\approx 1.001$), it is still necessary to treat the X and Y directions independently, as the results from our chosen path through the first Brillouin zone confirm. 

Figure~\ref{fig:bands_and_wfns}(b) and (c) show close-up views of the frontier HOMO and LUMO bands, calculated with PBE. Both are weakly dispersive, with measured bandwidths between 3 and 4~meV, defining an indirect band gap of 1.08~eV. To confirm that the frontier active space is well separated from surrounding states, panel~(d) displays a wide-window band structure, illustrating that the HOMO is isolated by over 1~eV from both the unoccupied LUMO/LUMO+1 manifold above and the occupied states below. Panel~(e) details a deeper valence region, where labels (i)--(iii) mark the most dispersive bands located well below the active manifold (between $-5.0$~eV and $-3.4$~eV). The HOMO band remains essentially isotropic, while anisotropy in the LUMO band reflects the intrinsic molecular $C_2$ symmetry. This orbital character is also illustrated by the $\Gamma$-point Bloch orbitals for the thin film [panel~(f)] and an isolated molecule [panel~(g)]; the HOMO has negligible amplitude near the two central hydrogens, whereas the LUMO exhibits significant weight in that region. By contrast, the deeper bands (i)--(iii) are more delocalized and extend onto neighboring molecules, consistent with their greater intermolecular hopping. These observations motivate the anisotropic nearest-neighbor tight-binding description which we fit band-by-band in the next section.

At liquid nitrogen temperatures ($k_B T \sim 6.6$~meV), thermal motion permits fluctuations across the central hydrogen tautomer orientations.~\cite{NC_switch,Naph_switch2} To quantify this effect, we evaluated the relative single-molecule ground-state energies ($\Delta E_{\text{mol}}$) for four tautomeric ordering configurations in the $2\times2$ supercell: $(0,0)$, $(\pi,0)$, $(0,\pi)$, and $(\pi,\pi)$ (see Fig.~\ref{S-fig:Anisotropy} in the SI). The resulting $77$~K Boltzmann populations show that the fully aligned $(0,0)$ state is the global minimum ($48.7\%$ weight) and the remaining states lie within $5\text{--}10$~meV/molecule, remaining thermally accessible. Because the HOMO orbital density resides away from the central hydrogens, its dispersion remains isotropic across all four configurations. Thus, thermal averaging yields an effectively $C_4$-symmetric spectral profile near $E_F$ without altering the predicted bandwidth ($\sim 3\text{--}4$~meV). Full energetic details and Boltzmann statistics are provided in SI Section~\ref{S-sec:struct_char}. 

\subsection{Effective electronic parameters and Coulomb interactions}

Based on the calculated electronic band dispersion and orbitals for the molecular network of H$_2$Nc, we derive the effective hopping parameters and Coulomb interactions, which serve as the fundamental competing energies in correlated electron systems. We express each single-particle DFT band with a minimal anisotropic tight-binding model to obtain the hopping parameters. For short-ranged Coulomb interactions, we take maximally-localized Wannier functions (MLWF)~\cite{wannier90} for the HOMO band and evaluate the relevant interaction integrals. This approach enables us to treat both the kinetic and interaction energy scales on an equal footing and compare them directly.

\begin{figure*}
\includegraphics[width=.8\textwidth]{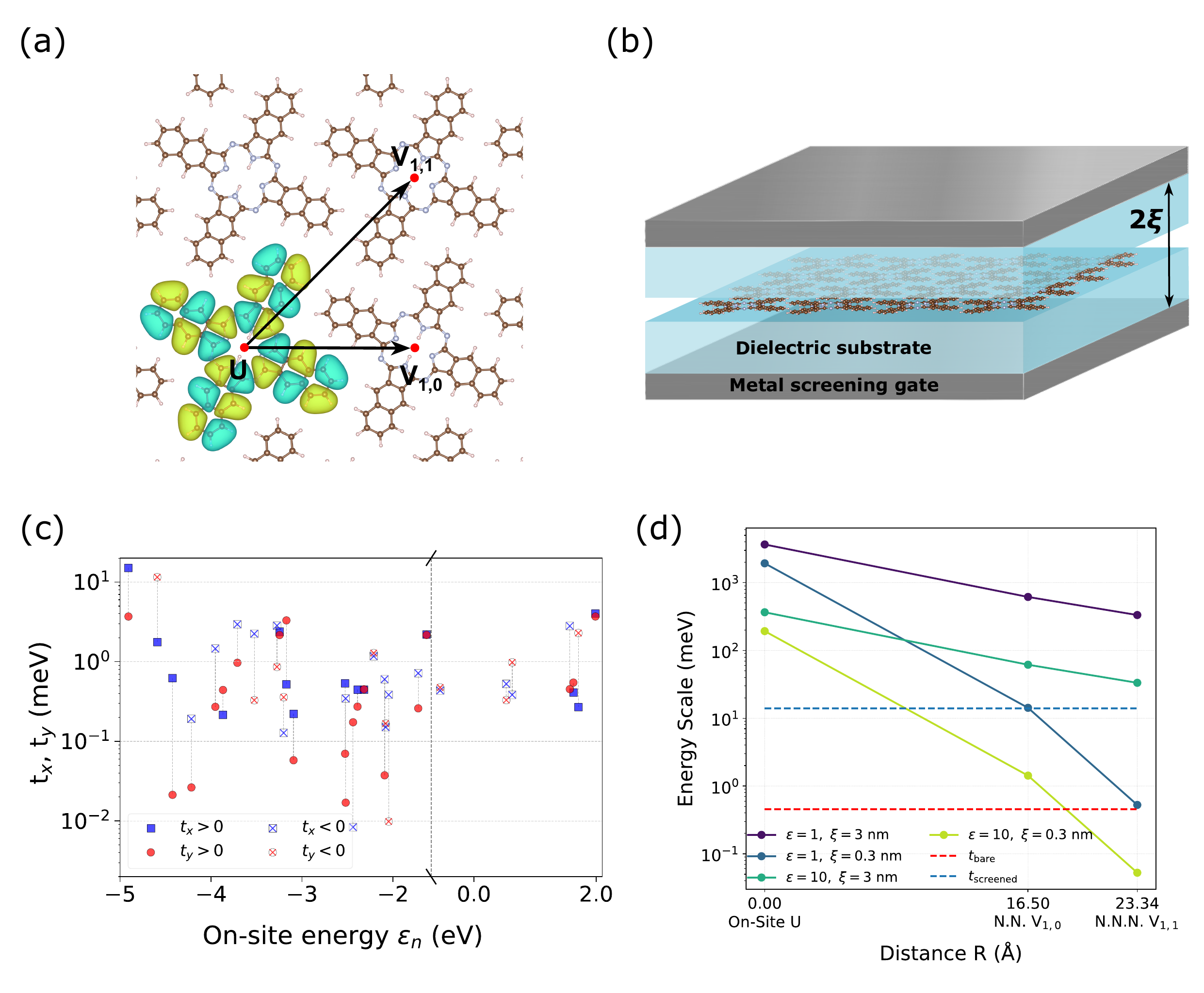}
\caption{(a) Maximally-localized Wannier orbital for HOMO in the free-standing H$_2$Nc monolayer. $U$ represents the on-site Coulomb interaction, and sites labeled $V_{1,0}$ and $V_{1,1}$ indicate the nearest- and next-nearest-neighbor vectors used to extract short-range Coulomb interactions $V_{ij}$. (b) Schematic of the model geometry and parameters entering the screening kernel (see text and Eqn.~(\ref{eq:kernel})). (c) Tight-binding parameters $\varepsilon_n$, $t_{x,n}$, and $t_{y,n}$, for the free-standing 2D thin film, for all bands in an energy window around $E_F$ (shaded/open markers distinguish positive/negative hoppings). (d) Computed on-site $U$ and short-range intersite $V_{ij}$ for several choices of the effective dielectric $\varepsilon$ and screening length $\xi$.}
\label{fig:Hubbard_U}
\end{figure*}

For each band $n$, we fit the dispersion to an anisotropic nearest-neighbor model:
\begin{equation} 
H_n=\sum_\mathbf{k}\Big[\varepsilon_n+2t_{x,n}\cos(k_x a_x)+2t_{y,n}\cos(k_y a_y)\Big]c^\dag_\mathbf{k} c_\mathbf{k},
\label{eq:Hamiltonian}
\end{equation}
extracting the on-site energy $\varepsilon_n$ and directional hoppings $t_{x,n}$, $t_{y,n}$. These fits were performed by least-squares regression on the DFT eigenvalues along the high-symmetry lines shown in Fig.~\ref{fig:bands_and_wfns}(a), yielding high fidelity per band. Fig.~\ref{fig:Hubbard_U}(c) reports the fitted $\varepsilon_n$, $t_{x,n}$ and $t_{y,n}$ for all bands in the chosen energy window satisfying $R^2 \geq 0.98$, providing a compact parameterization for subsequent model and transport studies. Table~\ref{S-tab:tb_pbe_vs_hse_all} tabulates the complete set, with further discussion on the validity of the nearest-neighbor model in SI Section~\ref{S-sec:nnn_model}. The frontier HOMO and LUMO bands, which set every quantity used in this work, are both fitted at $R^2 = 0.997$ and $0.996$, respectively. As the above calculations and analysis were implemented using the PBE functional, we further utilize the HSE06 hybrid functional~\cite{HSE06,HSE06_2} at the PBE-relaxed geometry (Methods) and observed the band gap increased from 1.08~eV to 1.37~eV; both values underestimate the $1.80 \pm 0.04$~eV single-particle gap measured for H$_2$Nc/Au(111) by orbital-mediated tunneling spectroscopy~\cite{Naph2}, itself a lower bound on the free-standing gap because image-charge screening by the metal is absent from our calculation, while PBE more closely matches the $\approx 1$~eV HOMO--LUMO separation resolved by ARPES for the adsorbed H$_2$Nc/Ag(100) system discussed in Section~\ref{sec:exp}. HSE06 also changes the frontier hoppings by no more than $\sim 25\%$, leaving the sign of every frontier orbital hopping, the near-isotropy of the HOMO band, the anisotropy of the LUMO band, and the correlation regime unchanged. Full comparison, including the 16-band parameter set for both functionals, is given in SI Section~\ref{S-sec:hse_comparison}. We therefore employ the PBE-calculated value throughout the remaining analysis of the frontier orbital hopping.

To probe the short-range Coulomb interactions entering the Hubbard model, we follow the method described in Refs.~\cite{U_Method1,U_Method2} to compute the Coulomb matrix elements for the HOMO Wannier function $w(\mathbf{r})$. Defining the Wannier density and its Fourier transform as:
\begin{equation}
\rho(\mathbf{r})=|w(\mathbf{r})|^2, \qquad \rho_{\mathbf{q}}=\int d^2\mathbf{r}\,\rho(\mathbf{r})e^{-i\mathbf{q}\cdot\mathbf{r}},
\end{equation}
the interaction at lattice separation $\mathbf{R}$ is given by:
\begin{equation}
U(\mathbf{R})=\frac{1}{(2\pi)^2}\int d^2\mathbf{q}\,V(\mathbf{q})|\rho(\mathbf{q})|^2 e^{i\mathbf{q}\cdot\mathbf{R}}.
\end{equation}
We evaluate this numerically over an auxiliary real-space cell of area $A$, taken large enough that periodic images of $\rho(\mathbf{r})$ do not interact, as the discrete sum:
\begin{equation}
U_{ij}\equiv U(\mathbf{R}_{ij})=\frac{1}{A}\sum_{\mathbf{q}}V(\mathbf{q})|\rho(\mathbf{q})|^2 e^{i\mathbf{q}\cdot\mathbf{R}_{ij}},
\end{equation}
where $\mathbf{R}_{ij}$ is the intersite vector from site $j$ to site $i$, and the on-site Hubbard $U$ corresponds to $U\equiv U_{ii}=U(\mathbf{0})$ (SI Section~\ref{S-sec:coulomb_method}). Environmental screening is incorporated via the interaction kernel:
\begin{equation}
V(\mathbf{q})=\frac{e^2}{2\varepsilon\varepsilon_0|\mathbf{q}|}\tanh(|\mathbf{q}|\xi),
\label{eq:kernel}
\end{equation}
where $\varepsilon$ is an effective, isotropic background dielectric constant for both the superstrate and substrate, and $\xi$ the distance to the top and bottom metallic screening gates.~\cite{Kernel1,Kernel2,Kernel3} %We treat $\varepsilon$ as a single scalar rather than a tensor: for a monolayer between two dissimilar isotropic media it should be read as the average of the substrate and superstrate responses, $\varepsilon \simeq (\varepsilon_{\mathrm{sub}} + \varepsilon_{\mathrm{sup}})/2$. 
Uniaxial anisotropy of the environment does not change the form of Eqn.~(\ref{eq:kernel}): it returns the same kernel with $\varepsilon \rightarrow \sqrt{\varepsilon_{\parallel}\varepsilon_{\perp}}$ and $\xi \rightarrow \xi\sqrt{\varepsilon_{\parallel}/\varepsilon_{\perp}}$ (SI Section~\ref{S-sec:coulomb_method}). As an example, for hBN this gives $\tilde{\varepsilon} \approx 5.1$ and a ${\sim}35\%$ rescaling of $\xi$, both well inside the ranges scanned in Fig.~\ref{fig:Hubbard_U}(d). Further details on the numerical implementation are provided in SI Section~\ref{S-sec:coulomb_method}.

Figure~\ref{fig:Hubbard_U} summarizes the calculated hopping parameters and screened Coulomb interactions. Figure~\ref{fig:Hubbard_U}(a) displays the HOMO MLWF isosurface used to evaluate $\rho(\mathbf{q})$, defining the on-site $U$ and short-range intersite interactions $V_{1,0}$ and $V_{1,1}$. Panel (b) depicts the dual-gated dielectric geometry modeled by Eqn.~(\ref{eq:kernel}), which reflects an ideal way to experimentally control the local Coulomb interaction strength. In the free-standing monolayer limit ($\varepsilon=1$, $\xi\rightarrow\infty$), we obtain a bare on-site value of $U_{\text{bare}}=3.99\text{ eV}$, with short-range repulsions $V_{1,0}=0.930\text{ eV}$ and $V_{1,1}=0.634\text{ eV}$. As shown in Fig.~\ref{fig:Hubbard_U}(d), dielectric screening systematically renormalizes these interaction scales downward. 

We emphasize that $U_{\mathrm{bare}} = 3.99$~eV is an unscreened upper bound: the bare Wannier integral neglects internal polarizability from the remaining bands of the molecular monolayer. Benchmark constrained random phase approximation (cRPA) calculations on molecular crystals and organic conductors with comparable $\sim 1$~eV frontier gaps reduce the on-site interaction to below one quarter of its bare value.\cite{cRPA1,cRPA2} Applying such a reduction to the free-standing monolayer gives $U_{\mathrm{cRPA}} \sim 1$~eV against a HOMO $t_{bare} \sim 0.45$~meV, so $U/t \gtrsim 10^3$ and the free-standing film remains deep in the localized limit irrespective of internal screening. The adsorbed case is treated separately in Section~\ref{sec:substrate_theory}, where the dominant renormalization of $U$ and $t$ arises from metallic screening and hybridization rather than from intramolecular polarizability; the $U/t_{\mathrm{screened}}$ values quoted there should likewise be read as upper bounds.

To quantify the correlation strength, these interaction scales must be compared with the kinetic hopping parameters. In Fig.~\ref{fig:Hubbard_U}(d) we contrast the free-standing hopping $t_{\mathrm{bare}} \approx 0.45$~meV with the substrate-enhanced hopping $t_{\mathrm{screened}} \approx 14.0$~meV derived for noble-metal-supported films in Section~\ref{sec:substrate_theory}. In each case the ratio is quoted across the full range of environmental screening spanned in Fig.~\ref{fig:Hubbard_U}(d), from the unscreened limit ($U = U_{\mathrm{bare}} = 3.99$~eV) to the most strongly gated geometry ($U \approx 200$~meV at $\varepsilon=10$, $\xi=0.3$~nm). For the free-standing film this yields $U/t_{\mathrm{bare}} \sim 400$--$8900$, placing the unsupported monolayer in an extreme localized limit. Incorporating substrate hybridization reduces the ratio to $U/t_{\mathrm{screened}} \sim 14$--$285$. %Since $U$ is an unscreened upper bound at every point in this range, the quoted ratios are themselves upper bounds. 
The internal-polarizability-screening reduction discussed above would carry the strongly screened end into the single digits, taking it across the square-lattice Mott boundary relevant to cuprates ($U/t \sim 8-12$)~\cite{Mott_Ratio} rather than leaving the accessible range entirely on the localized side. Comparing the four kinds of dielectric substrate setups as shown in Fig.~\ref{fig:Hubbard_U}(d), a larger dielectric constant $\varepsilon$ and a smaller substrate height $\xi$ are shown to bring down the energy scale of the interactions. For the smaller substrate height of $\xi = 0.3$~nm as shown in Fig.~\ref{fig:Hubbard_U}(d), the nearest- and next-nearest-neighbor interactions $V_{1,0}$ and $V_{1,1}$ are suppressed more strongly compared to the large substrate height of  $\xi = 3$~nm. Thus, we demonstrate that dielectric environment and substrate hybridization together renormalize $U/t$ by more than two orders of magnitude within a single molecular film, covering from an extreme localized regime down to and across the vicinity of the square-lattice Mott physics relevant to cuprates. 

\begin{figure*}
\includegraphics[width=.8
\textwidth]{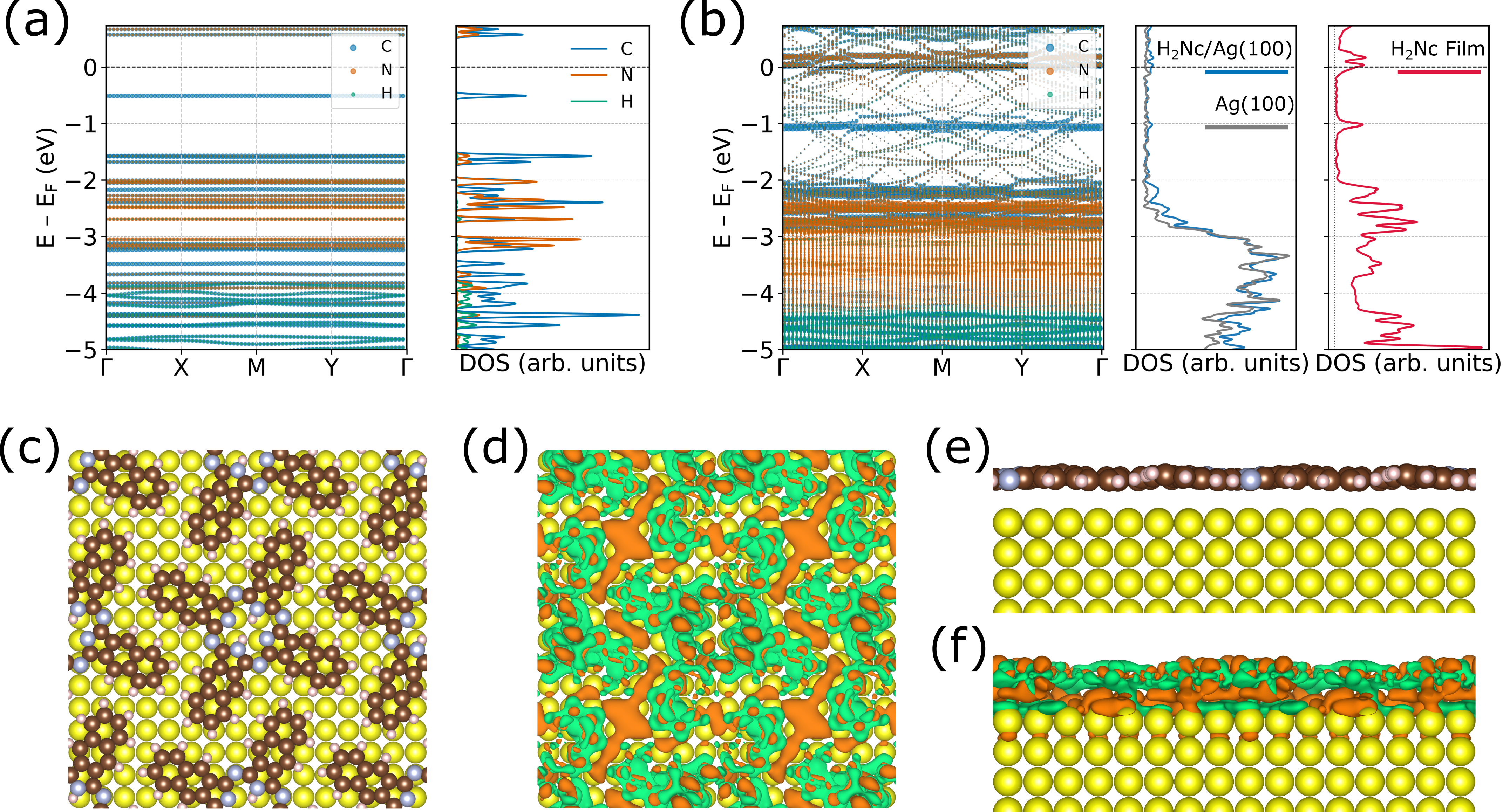}    
\caption{(a) Projected band structure and density of states (DOS) of the free-standing H$_2$Nc monolayer. (b) Projected band structure of the monolayer adsorbed on Ag(100), together with two complementary DOS panels: the middle panel compares the DOS of the adsorbed system (top Ag(100) layer + H$_2$Nc, blue) against the reference DOS of the bare Ag(100) top layer (grey), and the right panel shows the same adsorbed-system DOS projected only onto the molecular (C, N, H) orbitals (red), isolating the H$_2$Nc-derived spectral weight from the substrate background. A 20~meV Gaussian kernel is applied in post-processing to approximate the experimental energy resolution. (c,e) Top and side views of the relaxed H$_2$Nc/Ag(100) geometry. Further structure characterization can be found in Section~\ref{S-sec:struct_char} of the SI. (d,f) Top and side views of the bonding charge density difference, with isosurfaces shown at $3\times10^{-4}$ e$^-$/\AA$^3$. Orange represents electron accumulation and green shows depletion.}
\label{fig:4_5}
\end{figure*}

\subsection{Substrate-mediated band modulation in H$_2$Nc monolayers}
\label{sec:substrate_theory}
While the previous section models the system under an idealized experimental realization of $U/t$ control based on the dual-gated dielectric geometry, a simpler and much more widely practiced experimental setup would be to directly adsorb molecules onto metal substrates. This case would intuitively mimic the dielectric geometry at the thin-limit; but complications due to direct molecule-substrate electronic hybridization can arise depending on the interaction strength. To answer this matter-of-principle feasibility question, we now examine how these parameters are modified upon adsorption on Ag(100), as summarized in Fig.~\ref{fig:4_5}. Panels (a) and (b) compare the projected band structures and density of states (DOS) for the isolated monolayer and the adsorbed film. In contrast to the free-standing case, where the DOS vanishes at the Fermi level, adsorption produces partially filled, dispersive states with strong molecular character. These states appear as broadened molecular resonances that hybridize with Ag substrate states, inducing both energy shifts and broadening. The reference DOS of the bare Ag(100) slab, shown in the middle panel of Fig.~\ref{fig:4_5}(b), differs noticeably from the combined system, confirming that the electronic structure of the interface is distinct from both isolated constituents. 

Despite this hybridization, the molecular frontier manifold remains spectrally resolved after adsorption. The DOS projected onto molecular (C, N, H) orbitals, isolated in the rightmost panel of Fig.~\ref{fig:4_5}(b), retains a well-defined HOMO-derived peak that maps one-to-one onto the corresponding band of the free-standing film [Fig.~\ref{fig:4_5}(a)], rather than dissolving into a featureless continuum. Substrate hybridization enters as a band broadening and a rigid energy shift, both set by the molecule--metal coupling and well separated in scale from the $\sim 1$~eV gaps isolating the HOMO from the deeper occupied manifold and from the LUMO; adsorption therefore renormalizes the HOMO-derived band without closing these gaps or admixing additional orbitals into the active space at $E_F$. The single-band mapping used here thus applies to the HOMO manifold, the state probed in ARPES. The effective hopping $t_{\text{screened}} \approx 14.0$~meV quoted in Fig.~\ref{fig:Hubbard_U}(d) is obtained from the bandwidth of a molecular-projected, energy-broadened spectral function via the isotropic nearest-neighbor relation $W=8t$ (SI Section~\ref{S-sec:t_screened}), a $\sim 30$-fold enhancement over $t_{\text{bare}} \approx 0.45$~meV.

Next, we investigate interfacial charge transfer at the atomic scale. Figs.~\ref{fig:4_5}(c) and (e) illustrate the relaxed adsorption geometry on Ag(100). To visualize interfacial charge rearrangement, Figs.~\ref{fig:4_5}(d) and (f) plot the bonding charge-density difference $\Delta\rho(\mathbf{r})=\rho_{\text{combined}}(\mathbf{r})-\rho_{\text{slab}}(\mathbf{r})-\rho_{\text{molecule}}(\mathbf{r})$. Accumulation (orange) and depletion (green) isosurfaces (plotted at $\Delta\rho=3\times 10^{-4}\,e^-/\text{\AA}^3$) highlight local charge redistribution. Charge accumulation is concentrated in the interstitial region between the substrate and macrocycle, whereas the peripheral molecular arms show mild depletion—a pattern characteristic of physisorbed $\pi$-conjugated networks on noble metals~\cite{isoSmall,isoBig}.

To rigorously quantify this interfacial charge transfer, we computed the plane-averaged charge density difference, $\Delta\bar{\rho}(z)$, and its cumulative integral, $\Delta Q(z)$ (see SI Section~\ref{S-sec:charge_transfer}). Integration along the surface normal reveals a pronounced interfacial charge redistribution rather than a significant global transfer. Electron density strongly accumulates at the immediate H$_2$Nc interface, corresponding to a local peak of $\approx 0.22\,e^-$ in $\Delta Q(z)$. This interfacial gain confirms the partial filling of the H$_2$Nc LUMO via substrate back-donation after adsorption. However, this accumulation is accompanied by strong intramolecular polarization; a localized depletion region directly above the interface removes $\approx 0.12\,e^-$ from the upper molecular framework. To determine the final net molecular charge, we employed DDEC6 iterative Stockholder partitioning~\cite{DDEC6_1,DDEC6_2}, which is highly robust against boundary artifacts common at metal-organic interfaces. DDEC6 yields a net gain of merely $+0.10\,e^-$ per molecule, accurately reflecting the cancellation between the local LUMO accumulation and the internal $\sigma$-depletion. For comparison with the wider literature, where Bader charges~\cite{Bader1,Bader2,Bader3,Bader4} are commonly quoted for this class of interface, zero-flux topological partitioning of the same converged density gives $0.82\,e^-$. We attribute the larger value to the assignment of diffuse metallic surface density to the molecular basin rather than to a larger physical transfer.

We stress that these measures answer different questions: $\Delta Q(z)$ reports the local accumulation on the molecular side of the interface, while the DDEC6 value reports the net charge of the molecule. Neither is a rigorous orbital occupancy. Together, the control over dielectric screening demonstrated in Fig.~\ref{fig:Hubbard_U}(d), the substrate-induced enhancement of $t_{\mathrm{screened}}$, and the local interfacial charge redistribution quantified above establish a broad accessible parameter space for H$_2$Nc superlattices.

\begin{figure}[!tbh]
\includegraphics[width=.5\textwidth]{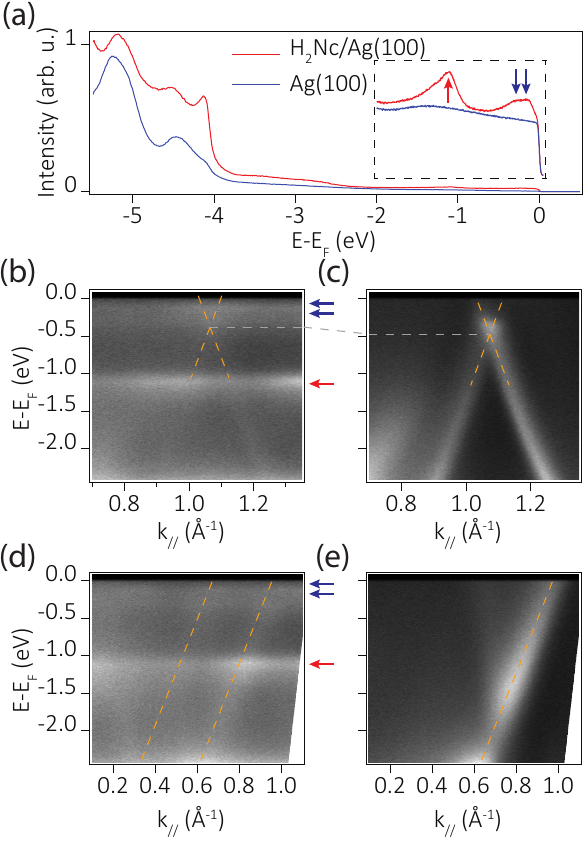}
\caption{Electronic structure of H$_2$Nc molecular superlattice on Ag(100). (a) Comparison of UPS spectra of pristine Ag(100) surface (blue curve) and H$_2$Nc monolayer grown on top (red curve). Inset shows high-resolution zoom-in view of the low-energy electronic states within binding energy between -2.00 eV and 0.05 eV. Red arrow denotes the HOMO band (also in (b) and (d)), and blue arrows denote the two LUMO bands (also in (b) and (d)). Energy-momentum cuts through $\Gamma-$X direction of the Ag(100) Brillouin zone on (b) monolayer H$_2$Nc/Ag(100) and (c) pristine Ag(100) surfaces. Yellow dashed lines denote the conduction bands of Ag(100). The grey dashed lines in (b) and (c) imply an upward conduction band shift of 90 meV due to interfacial charge transfer when H$_2$Nc is adsorbed. Energy-momentum cuts through $\Gamma-$M direction of the Ag(100) Brillouin zone on (d) monolayer H$_2$Nc/Ag(100) and (e) pristine Ag(100) surfaces. (d) shows a replica of the conduction band due to the superlattice potential of H$_2$Nc. All measurements are at 82 K.}
\label{fig:Ag100}
\end{figure}

\begin{figure}[!tbh]
\includegraphics[width=.5\textwidth]{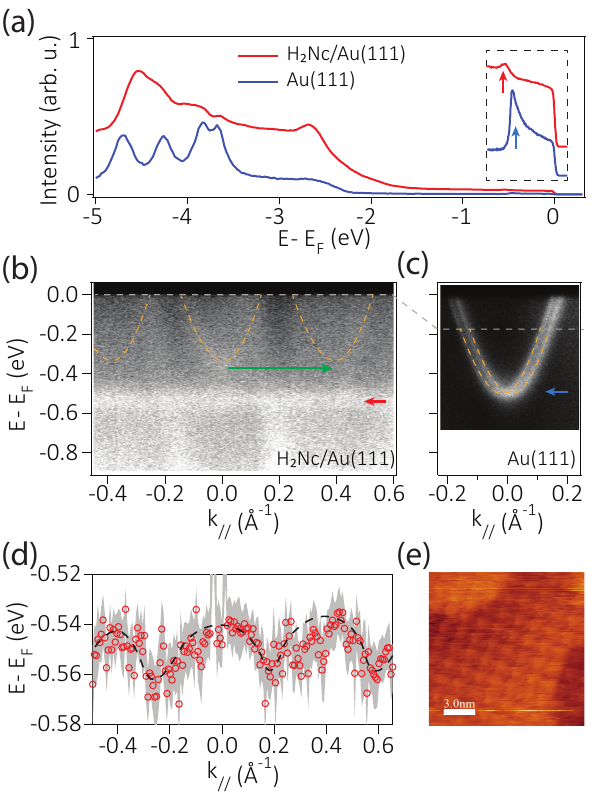}
\caption{Electronic structure of H$_2$Nc molecular superlattice on Au(111). (a) Comparison of UPS spectra of pristine Au(111) surface (blue curve) and H$_2$Nc monolayer (red curve). Inset shows high-resolution zoom-in view of the low-energy electronic states within energy ranging from -0.75 eV to 0.05 eV. Red arrow denotes the HOMO band (also in (b)), and blue arrow denotes the Shockley surface state band bottom (also in (c)). Energy-momentum cuts through $\Gamma$ on (b) monolayer H$_2$Nc/Au(111) and (c) pristine Au(111) surface. Yellow dashed lines denote the Shockley surface state of Au(111). The grey dashed lines in (b) and (c) imply an upward surface state shift of 200 meV due to charge transfer when H$_2$Nc is adsorbed. The green arrow indicates the momentum scattering vector of the Au surface state resulting from the H$_2$Nc superlattice. (d) Energy dispersion of the HOMO band fitted from the energy dispersive curves (EDCs) of the ARPES spectra using a Gaussian peak. The red dots denote the Gaussian peak positions of the EDCs and the grey bars are the standard error of Gaussian peak fit plus thermal broadening. The black dashed line is a guide to the eyes of the periodic dispersion. (e) Topographic image of local patches of H$_2$Nc molecular superlattice on Au(111). ARPES and STM are measured at 82 K and 295 K respectively.}
\label{fig:Au111}
\end{figure}

\subsection{Experimental observation of substrate-tunable electronic structure}
\label{sec:exp}

While direct gating through a dielectric ``spacer layer'' is the ideal geometry for the device control of $U/t$, assembling molecules directly on bare metal substrates is much easier experimentally and can offer a direct experimental test at the thin dielectric layer limit. The electronic structure modification due to substrate-molecule interaction is investigated experimentally with angle-resolved photoemission spectroscopy (ARPES) with the help of STM to identify the domain sizes. To directly compare with the calculation, monolayer H$_2$Nc is grown on a sputter-annealed Ag(100) surface via a home-built molecular beam epitaxy (MBE) system. Fig.~\ref{fig:Ag100}(a) shows the low energy electronic density of states (DOS) of pristine Ag(100) and H$_2$Nc/Ag(100) using ultra-violet photoemission spectroscopy (UPS). Compared with the theoretical results in the right panel of Fig.~\ref{fig:4_5}(b), the energy difference between the HOMO (red arrow) and the LUMOs (blue arrows) is about 1 eV, which is almost the same as the first-principles H$_2$Nc/Ag(100) calculation. The broad asymmetric HOMO peak indicates strong electronic mixing between the molecular layer and the Ag substrate which is consistent with the band hybridization between Ag(100) and H$_2$Nc calculated in Fig.~\ref{fig:4_5}(b). 
%However, the experimental LUMOs and HOMO have about 250 meV deeper binding energies compared to the \textit{ab initio} results. This binding energy discrepancy could possibly result from the electron correlations in the system. 
The comparison between the energy-momentum spectra of H$_2$Nc/Ag(100) (Fig.~\ref{fig:Ag100}(b)) and Ag(100) (Fig.~\ref{fig:Ag100}(c)) illustrates an $\sim 90$~meV upward shift in energy of the Ag(100) conduction band along the $\Gamma-$X direction, consistent with the interfacial charge redistribution obtained from the plane-averaged density difference in the previous section, although the LUMO-derived filling inferred from the ARPES data is larger than the net molecular charge obtained from partitioning. This is not in conflict, as photoemission constrains the interfacial level alignment, which is set by the local dipole, whereas the partitioned charge reports the net occupancy of the molecular basin, and the two are decoupled by the intramolecular $\sigma$-depletion identified above. The superlattice potential exerted by the H$_2$Nc network on the Ag(100) surface is faint but identifiable and visible through the replica of the Ag conduction band in momentum space along the $\Gamma-$M direction (Fig.~\ref{fig:Ag100}(d)) in contrast to the bare Ag(100) (Fig.~\ref{fig:Ag100}(e)). However, the single-band tight-binding fit of the HOMO band is less well-defined (see Section~\ref{S-sec:Ag100_fitting} of the SI) due to its strong hybridization with the Ag(100) conduction bands implied by both the large FWHM of 350~meV in Fig.~\ref{fig:Ag100}(a) inset and theory results in Fig.~\ref{fig:4_5}(b).

To circumvent the complication due to strong substrate-molecule hybridization on Ag(100) surface, and to also demonstrate the substrate tunability, monolayer H$_2$Nc is grown on a sputter-annealed Au(111) surface using the same method. After growth, the topography of the molecule layer is measured by STM, which shows patches of H$_2$Nc monolayers with a typical size of $20\times20$ nm$^2$ (Fig.~\ref{fig:Au111}(e)).  Figure~\ref{fig:Au111}(a) shows the low energy electronic density of states (DOS) of pristine Au(111) and H$_2$Nc/Au(111) UPS. Compared to H$_2$Nc/Ag(100) in Fig.~\ref{fig:Ag100}(a), the HOMO peak of H$_2$Nc/Au(111) in the UPS is a sharp and symmetric peak with a smaller 100~meV FWHM at $\sim0.55$~eV binding energy. This indicates less electronic mixing between the molecule layer and Au substrate in contrast to the case for Ag substrate~\cite{AgAuInteractionComparison1,AgAuInteractionComparison2}, which is further supported by the energy-momentum spectra shown in Fig.~\ref{fig:Au111}(b) and (c). The Shockley surface state (yellow dash) can still be resolved in the case of H$_2$Nc/Au(111) especially near $\Gamma$,~\cite{shockley1939surface} and is shifted upward by $\sim 200$~meV, consistent in sign with the interfacial charge redistribution obtained from the plane-averaged density difference in the previous section, which shows electron accumulation in the lower molecular $\pi$-lobes upon adsorption. We note that the calculation refers to Ag(100) and this measurement to Au(111), so the comparison is to the direction of the interfacial dipole rather than to its magnitude. As a result, the HOMO band of H$_2$Nc sits roughly 550 meV below the Fermi level (red arrow, also shown in Fig.~\ref{fig:Au111}(a) inset). 

The superlattice potential exerted by the H$_2$Nc network on the Au(111) surface is also visible through the faint Umklapp replicas of the Shockley state in momentum space, with the momentum scattering vector which is about 0.4 \AA$^{-1}$ (the green arrow in Fig.~\ref{fig:Au111}(b)) matching the reciprocal lattice vector of the H$_2$Nc assembly measured by STM which is $2\pi/1.7$ nm $\approx 0.37$ \AA$^{-1}$.~\cite{Naph,Naph2} The dispersion bandwidth of the HOMO band shows a periodic oscillation according to the molecular lattice and exhibits a bandwidth of about 50 meV, an order of magnitude larger than that of a free-standing H$_2$Nc film (for the repeatability of the dispersion, see Section~\ref{S-sec:Au111_repeatability} of the Supplementary Information (SI)). This qualitatively matches the expected electron kinetic energy enhancement due to more benign hybridization with metal conduction electrons in contrast to the experimental and theoretical results for the H$_2$Nc/Ag(100) interface in Fig.~\ref{fig:4_5}(b). Compared to previous molecular assembly studies using larger molecules, the H$_2$Nc lattice adopted here measures above the typical Fermi wavelength of the Shockley surface state electron gas, which avoided direct Umklapp band hybridization seen in other systems.~\cite{EngineeringQuantumStates2022} Finally, due to the tautomerization associated with the inner two hydrogen atoms' position fluctuations -- due to both thermal activation and quantum tunneling~\cite{Tautomerization1,Tautomerization2, Naph_switch2} -- the in-plane dispersion anisotropy is not expected in these measurements at liquid nitrogen temperatures. Such fluctuation effects, especially in the quantum regime, would require computational treatments beyond the Boltzmann-averaging discussed in SI Section~\ref{S-sec:struct_char} to fully capture.

\section*{Conclusion and Discussion} \label{sec:conclusions}
We used first-principles calculations and surface spectroscopy to quantify how intermolecular hopping, band dispersion, and Coulomb interactions in self-assembled H$_2$Nc monolayers are modified by metal substrates. Tight-binding fits of the free-standing film's electronic band structure reveal very small intrinsic hoppings and correspondingly large on-site Coulomb energies (unscreened $U$ up to $3.99$ eV). When the film is supported on a noble metal, metallic screening and hybridization enhance effective hopping roughly thirty-fold and reduce $U/t$, as confirmed by ARPES which shows a significantly broadened HOMO band compared to the thin film. Our calculations show that substrate choice, film-substrate spacing, and dielectric environment can tune $U/t$ across orders of magnitude, enabling exploration from the extreme localized limit to strongly correlated regimes with the same molecular film. 

More broadly, this tunability spans distinct classes of substrate. Weakly hybridized platforms such as hexagonal boron nitride or graphene/Ir(111) moir\'{e} templates~\cite{sub_decoupling} are expected to leave $t$ close to its free-standing value while providing only modest dielectric screening, keeping $U/t$ near the upper end of the free-standing range computed here ($U/t_{\mathrm{bare}} \sim 400$--$8900$). On the other hand, direct adsorption on a noble metal, as realized on Ag(100) and Au(111), drives $U/t$ down by more than an order of magnitude ($U/t_{\mathrm{screened}} \sim 14$--$285$, potentially into the single digits once additional internal screening is included) by simultaneously enhancing $t$ and screening $U$. %The lattice model these parameters define, and the device requirements for reaching it, are further discussed in SI Section~\ref{S-sec:simulation}. Furthermore, while microscopic hydrogen tautomerism breaks local fourfold symmetry in static calculations, thermal sampling of tautomer configurations at 77 K restores an effective macroscopic $C_4$ symmetry across the electronic structure without altering the underlying orbital dispersion or bandwidth.

%In a physical system, increased kinetic hopping due to substrate hybridization delocalizes the molecular Wannier function, reducing the local interaction strength and further suppressing $U/t$. By adjusting the substrate work function, film-substrate separation, or dielectric spacer layer, $t$ and $U$ can be systematically manipulated. While we focus here on noble metal adsorption, decoupled templates such as graphene/Ir(111) moiré superlattices~\cite{sub_decoupling} offer complementary routes to access more weakly hybridized limits, establishing substrate selection as a broad chemical control over effective model parameters.

This tunable range positions H$_2$Nc superlattices as a platform for the quantum simulation across much of the phase diagram of the extended square-lattice Hubbard model~\cite{Arovas2022}. In the deep localized limit ($U >> t$), the half-filled film maps onto the $S = 1/2$ square-lattice Heisenberg antiferromagnet, and at commensurate fractional fillings the intersite repulsions stabilize generalized Wigner crystals~\cite{Tong2025GWC}; in the strongly correlated regime ($U/t \sim 8-12$), the model hosts the physics associated with the cuprates: the antiferromagnetic Mott insulator and, upon doping, stripe/charge-density-wave order and unconventional $d$-wave superconductivity~\cite{LeeNagaosaWen2006,Scalapino2012,Zheng2017}. We emphasize that although no correlation-driven phases are observed in the present experiments, these are the established phases of the model that the demonstrated parameter control would make accessible in a decoupled, gated geometry. %The relevant magnetic scale, the superexchange $J = 4t^2/U$, reaches tens of Kelvin at the substrate-enhanced hoppings quantified here but collapses in a fully decoupled film, where $t$ reverts toward $t_{\mathrm{bare}}$; the practical target is therefore the intermediate-screening regime of a thin dielectric spacer, where $t$ remains enhanced while $U$ and the filling are gate-controlled. 
Realizing this same square-lattice Hubbard Hamiltonian has been an actively pursued goal in twisted GeX/SnX~\cite{Xu2024GeSnX} and twisted-bilayer C$_{568}$~\cite{U_Method2} moir\'e systems; the self-assembled molecular route provides a complementary, \textit{in situ} scalable form of parameter control. Our system can also reach a tunable $t_x/t_y$ beyond what isotropic systems can provide.\cite{U_Method2}  %Complementing these two isotropic tunable systems, our system can also realize a potential lattice-anisotropy knob, as realized in the LUMO-derived band with anisotropic hopping of $t_x/t_y = 1.59$. 
An important caveat qualifies this knob: the $C_2$ anisotropy originates from a single frozen hydrogen-tautomer configuration, and thermal sampling of tautomer configurations at 77~K restores an effective macroscopic $C_4$ symmetry. The static anisotropy therefore are realized only if the tautomers are ordered at lower temperature or through local manipulation, recasting tautomerism potentially into a switchable anisotropy control.

These results highlight the molecular assembly/solid state interface as an exciting platform to engineer electronic properties: on the one hand, the molecules can act upon the solid state surface via a superlattice potential, creating mini Brillouin zones akin to those in moiré engineered systems but in an \textit{in situ} scalable way; on the other hand, the metal substrate can interact strongly with the periodic molecular lattice, causing enhanced electron hopping across molecules, and can potentially be exploited to modify the optical and magnetic response of such molecular thin films. Our results also show that in the absence of a dielectric-floated gate control, too strong direct molecule-substrate interaction can break up the molecular band and complicate theoretical descriptions. Alternatively, new strategies involving both dielectric spacer layers and molecular thin films grown or stacked above metal substrates also offer an exciting route towards the ideal dual-gated dielectric geometry shown in Fig.~\ref{fig:Hubbard_U}. Excitingly, recent developments on \textit{in situ} growth of molecular thin films over van der Waals substrates demonstrated pronounced moiré superlattice effects on the optical properties of the substrates, validating this new route of molecular superlattice engineering to design, create, and control new quantum material properties.~\cite{chowdhury2025bright,chowdhury2025spectra}

\section*{Methods} \label{sec:develop}

\subsection*{Density Functional Theory Calculations}
In order to compute the desired electronic properties of the various structures studied, first-principles density functional theory (DFT) calculations were performed using the Vienna \textit{ab-initio} Simulation Package (VASP).~\cite{vasp1, vasp2} Projector-augmented wave (PAW) potentials~\cite{PAW} were used, and exchange–correlation effects were treated within the generalized gradient approximation (GGA) as parametrized by Perdew, Burke, and Ernzerhof (PBE).~\cite{PBE,GGA} Dispersion interactions between nearby molecules in the H$_2$Nc system and between the molecular layer and the substrate were accounted for using the DFT-D3 method of Grimme \textit{et al}. with Becke–Johnson damping.~\cite{vdW1, vdW2} This choice of van-der-Waals functionals results in good agreement of the lattice parameters with experimental values.~\cite{Au_111} Wannier downfolding is implemented using maximally localized Wannier functions with Wannier90.~\cite{wannier90}

\subsection*{Isolated H$_2$Nc Thin Film}

Calculations for the isolated H$_2$Nc monolayer were done using a plane-wave basis set of energy cutoff 480 eV and an $11\times11\times1$ Monkhorst-Pack (MP) $k$-point mesh. The structural relaxation of both atomic positions and lattice parameters was performed until residual forces on all atoms were less than $1$ meV/\AA. Vacuum space of 20 \AA\ was maintained in the $c$ direction in order to generate an isolated monolayer. Charge densities for the electronic band structure, density of states (DOS), and $\Gamma$-point Bloch wavefunctions were converged using Gaussian smearing of 20~meV (\texttt{ISMEAR}~$=0$, \texttt{SIGMA}~$=0.02$). Because the free-standing monolayer is gapped by 1.08~eV, all occupations are integer at this smearing width and the resulting eigenvalues $E_n(\mathbf{k})$ carry no dependence on the smearing parameter. The reported HOMO and LUMO bandwidths, obtained directly from the discrete eigenvalue spectrum as $W = \max_{\mathbf{k}} E(\mathbf{k}) - \min_{\mathbf{k}} E(\mathbf{k})$, are therefore independent of it. DOS curves were subsequently evaluated with the linear tetrahedron method with Bl\"ochl corrections (\texttt{ISMEAR}~$=-5$), which performs exact analytic integration over $\mathbf{k}$-space and introduces no artificial broadening of its own. The 20~meV Gaussian appearing in Figs.~\ref{fig:4_5}(a) and 3(b) is a separate post-processing convolution applied to these exact, tetrahedron-integrated spectra for display only, chosen to approximate the experimental energy resolution; it does not enter any reported bandwidth, hopping parameter, or energy. 

To assess the dependence of the band dispersion and the extracted tight-binding parameters on the exchange--correlation functional, the free-standing H$_2$Nc monolayer was additionally treated with the screened hybrid functional HSE06.~\cite{HSE06,HSE06_2} The hybrid charge density and orbitals were converged self-consistently on a $3 \times 3 \times 1$ Monkhorst--Pack mesh, corresponding to a $k$-point spacing of $0.02 \times 2\pi/\text{\AA}$, using the converged PBE orbitals as the starting guess. The geometry was held fixed at its PBE+D3-BJ relaxed values; no structural relaxation was performed at the hybrid level. 

\subsection*{Molecular Layer on Ag(100) Substrate}

DFT calculations were performed to find the lattice constant of bulk Ag(100) to be 4.07 \AA, which agrees well with the high-precision experimentally measured value of 4.08 \AA.~\cite{Aga} A compressive strain of about 1\% was then applied to the relaxed H$_2$Nc thin film in order to commensurately place it above a $4\times4$ slab of Ag(100) (five atomic layers thick, in-plane lattice parameters  $a=b=16.29$ \AA). The Ag(100) slab was placed at the center of a periodic cell with the H$_2$Nc layer 3 \AA\ above it, with 11 \AA\ of vacuum below the bottom Ag layer and 8 \AA\ of vacuum above the molecular film to fully decouple periodic images. The lowest-energy placement of the molecular layer on the substrate was found by translating over the two-dimensional space of relative in-plane displacements on a grid of spacing $0.125\,a_0 \approx 0.51$~\AA, and performing $\Gamma$-point relaxations of the $z$-components of the atomic positions within the molecular layer to a force convergence of 1 meV/\AA. The energy difference between the highest- and lowest-energy placements of the molecular layer was approximately 220~meV, highlighting the importance of this grid-search procedure. The lowest energy placement preserves the symmetric registry of the molecule on the fourfold Ag(100) surface net, so the only symmetry reduction in the adsorbed system is the intrinsic molecular $C_2$ of the free H$_2$Nc macrocycle. The resulting energy landscape (Fig.~\ref{S-fig:translations}) is smooth on the scale of the sampling grid.

After choosing the lowest energy placement, all degrees of freedom of the atoms within the molecular layer and top substrate layer were relaxed with a $3\times3\times1$ MP-mesh to a force convergence of 1 meV/\AA. Further structure characterization of the thin film with and without the Ag(100) substrate, as well as a visualization of the energy landscape of relative film-substrate translations (Fig.~\ref{S-fig:translations}), can be found in the Section~\ref{S-sec:struct_char} of the SI. 

Using the fully relaxed geometry, calculations were performed to obtain the band structure, density of states, and bonding charge density difference of the substrate and H$_2$Nc layer. Because the structure contains both a bulk metal and an insulating molecular layer, occupations near $E_F$ are fractional and the converged density depends in principle on the smearing width. Calculations repeated at Gaussian widths from 50 meV down to 2 meV showed no changes affecting the quantities reported here; those presented correspond to 20 meV. The molecular HOMO manifold lies $\approx 1$~eV below $E_F$ and is fully occupied at all widths considered.

\subsection*{Growth of Naphthalocyanine on Ag(100) and Au(111) Substrates}

The growth was carried out in a homemade ultrahigh vacuum (UHV) molecular beam epitaxy (MBE) system with a base pressure better than $1.5\times10^{-10}$ Torr. A single-crystal Ag(100)/Au(111) (roughness $<0.01$ $\mu$m, orientation accuracy $<0.1^\circ$, Princeton Scientific) was prepared in the UHV MBE chamber by repeated cycles of argon sputtering and subsequent annealing at $500^\circ$C until a clean surface was confirmed by reflective high-energy electron diffraction (RHEED) or Auger electron spectroscopy (AES). Purified H$_2$Nc in powder form was degassed at $390^\circ$C for 1 hour under UHV using a Knudsen cell (Rocksteady). It was evaporated at $390^\circ$C with the Ag(100)/Au(111) substrate kept at room temperature. The evaporation rate was about 0.2 layers per minute monitored by a quartz crystal microbalance (Inficon).

\subsection*{Structural and Spectroscopic Characterization}

The structure of the film was characterized \textit{in situ} by a homemade scanning tunneling microscope (STM) using a Pt-Ir tip in constant-current mode at room temperature. Angle-resolved photoemission spectroscopy was carried out \textit{in situ} in a homemade system under UHV about $4\times10^{-11}$ Torr at 82 K. ARPES data were collected using the MB Scientific A-1 analyzer with an energy resolution of $15$ meV, determined by
fitting a Gaussian-convolved Fermi function. Monochromatized 21.2 eV VUV light was generated by a helium discharge lamp (SPECS UVS 300) delivered to the same surface via an ellipsoidal focusing capillary. The momentum resolution is estimated as $\sim 0.015$~\AA$^{-1}$ contributed from the $20\times20$ nm$^2$ domain size and the $\sim0.3^{\circ}$ instrumental angular resolution.

\section*{Supplementary Information} \label{sec:supplementary}

\textbf{Supporting Information Available:} Relaxed geometries of the free-standing and Ag(100)-supported H$_2$Nc monolayer, central-hydrogen tautomer configurations with their relative energies and 77~K Boltzmann populations, and the potential energy surface for lateral registry of the monolayer on Ag(100); comparison of PBE and HSE06 fundamental gaps and frontier hopping parameters, together with the complete 16-band tight-binding parameter set for both functionals; assessment of the nearest-neighbor tight-binding model and extended next-nearest-neighbor fits for the two bands with reduced fit quality; discussion on the parity of the $\Gamma$-point Bloch orbitals in the film relative to the isolated molecule; numerical method for evaluating the screened Coulomb integrals and their dependence on the effective dielectric constant, screening gate distance, and supercell padding; extraction of the effective screened hopping $t_{\mathrm{screened}}$ from a molecular-projected spectral function; interfacial charge redistribution from the plane-averaged charge-density difference and comparison of three charge-partitioning schemes; lattice models and device requirements for quantum simulation; 
repeatability of the fitted H$_2$Nc/Au(111) HOMO dispersion away from normal emission; ARPES HOMO-band fitting for H$_2$Nc/Ag(100).
\vspace{0.5cm}
\section*{Acknowledgments} \label{sec:acknowledgements}

We thank Sam Trickey for the insightful discussion. We thank Xiaofan Jia and Nilay Hazari for H$_2$Nc powder purification. This work is supported by the Center for Molecular Magnetic Quantum Materials, an Energy Frontier Research Center funded by the U.S. Department of Energy, Office of Science, Basic Energy Sciences under Award no. DE-SC0019330. Computations were done using the utilities of the University of Florida Research Computing. Y. He and Z. Kang acknowledge the support of the Office of Naval Research (ONR) under grant No. N0014-23-1-2018. E. Altman acknowledges the support of the National Science Foundation through grant number CHE-2203589. Z. Zhu acknowledges a startup fund at Boston College.

%---------------------------------------------------------------------
% BIBLIOGRAPHY
%
% Only ONE \bibliography is allowed per document, so supplementary.tex's
% \bibliographystyle{apsrev4-1} + \bibliography{main} have been deleted.
% BibTeX collects citations from the whole file, so the 13 SI-only references
% are included in this list too, numbered after the 78 main-text ones (91
% total). SI citations therefore no longer restart at 1 as they did in the
% standalone SI.
%
% If you would rather the SI carry its own separately numbered reference
% list, that needs the bibunits package -- ask and I'll write that version.
%---------------------------------------------------------------------
\bibliographystyle{achemso}
\bibliography{main}

%=====================================================================
%  SUPPORTING INFORMATION
%=====================================================================
\clearpage
\onecolumngrid
\makeatletter
\let\subsection\subsection@preprintsty
\let\subsubsection\subsubsection@preprintsty
\renewcommand{\baselinestretch}{1.5}
\input{aps12pt4-1.rtx}
\makeatother
\newgeometry{margin=1in}
\normalsize

\setcounter{figure}{0}
\setcounter{table}{0}
\setcounter{equation}{0}
\setcounter{subsection}{0}
\setcounter{subsubsection}{0}
\renewcommand{\thefigure}{S\arabic{figure}}
\renewcommand{\thetable}{S\arabic{table}}
\renewcommand{\theequation}{S\arabic{equation}}
\setcounter{secnumdepth}{2}

\begin{center}
{\Large \textbf{Supporting Information}}\\%[1em]
{\large Self-Assembled Naphthalocyanine Molecular Lattices as a Platform for Tunable Quantum Superlattices}\\[0.5em]
Adrian Bahri, Zhibo Kang, Ziyan Zhu, Eric I. Altman, Yu He, and Chunjing Jia
\end{center}

\label{sec:extra} 

\subsection{Structure Characterization}
\label{sec:struct_char}

\begin{figure}[h]
    \includegraphics[width=.4\textwidth]{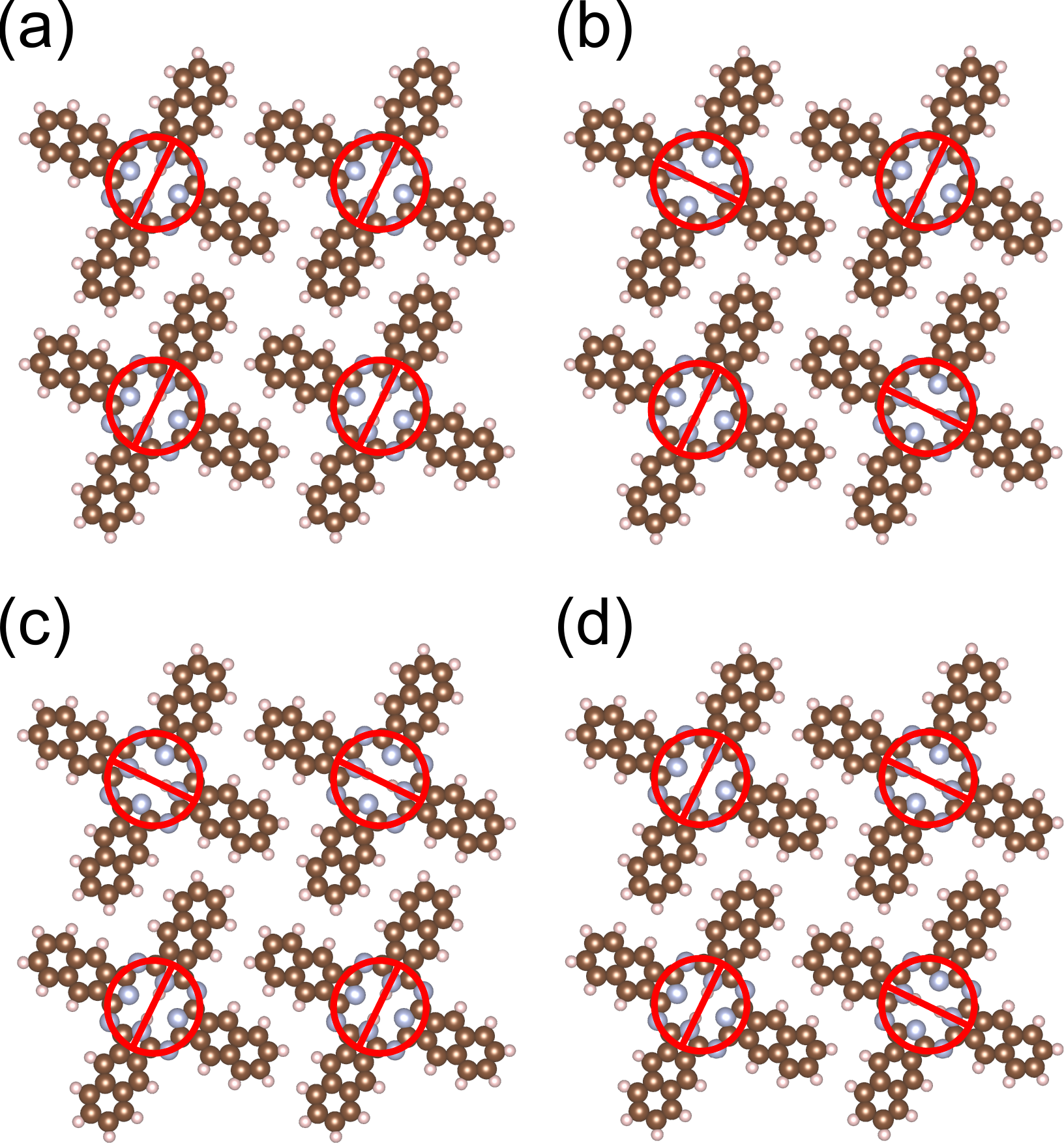}  
    \caption{Different molecular configurations studied by density functional theory. We refer to them as (a): $(0,0)$, (b): $(\pi,\pi)$, (c): $(\pi,0)$, (d): $(0,\pi)$ orders. Red circles over the macrocycles with inscribed lines emphasize hydrogen atom orientations. The $(0,0)$ order was found to be the global energy minimum, with the remaining configurations lying $5.45$--$9.81$~meV/molecule higher in energy (Table~\ref{tab:tautomers}).}
    \label{fig:Anisotropy}
\end{figure}

Exploratory DFT calculations were carried out to probe the orientations of the central hydrogen atoms within the H$_2$Nc macrocycle in the ground state of the thin film. Fig.~\ref{fig:Anisotropy} shows the four unique periodic orientations investigated. Atomic positions and lattice parameters of the H$_2$Nc thin film were relaxed in configuration (a) with calculation parameters detailed in the Methods. From this structure, configurations (b), (c), and (d) were generated, and atomic positions and lattice parameters were relaxed in each configuration to a convergence threshold of 1 meV/\AA. 

After relaxation, the resulting lattice parameters for the isolated H$_2$Nc layer were approximately 16.5~\AA\ in the $a$ and $b$ directions. Each molecule is approximately 2 nm wide and forms an angle of $25.8^\circ$ with the lattice vectors. The relaxed structure of the free-standing H$_2$Nc molecules is shown in Fig.~\ref{fig:bands_and_wfns} (a). 

After fully relaxing the thin film on the substrate, the mean height between the top layer of the substrate and the thin film was 3.03~\AA, and the range of atomic $z$-coordinates within the thin film was 0.56~\AA. These results resemble those of Wu \textit{et al.} for metal-free H$_2$Nc on Ag(111)~\cite{Naph} and those of Mehring \textit{et al.} for metal-free H$_2$Nc on Au(100).~\cite{Au100}

It is worth noting that Mehring \textit{et al.} grew the H$_2$Nc layer with an angle $\theta=0^\circ$ with respect to the substrate, meaning the lobes of each molecule were aligned parallel to the lattice vectors of the substrate, while in this work $\theta=25.8^\circ$. The structure was arranged as such for this work in order to minimize the size of the periodic unit cell and thus decrease the associated computational cost. Rotating the H$_2$Nc layer by a non-high-symmetry angle breaks commensurability with the Ag(100) lattice. To accurately study that system computationally would require either heavily straining the molecular film or constructing a large supercell to model the resulting Moir\'e pattern.

In the free-standing $\text{H}_2\text{Nc}$ monolayer, the two central hydrogen atoms break the local $C_4$ rotational symmetry of the isolated molecule down to $C_2$, orienting along one of two orthogonal molecular arms. In a $2\times2$ supercell containing four molecules, this binary orientational degree of freedom gives rise to four distinct long-range spatial ordering configurations:
\begin{itemize}
    \item \textbf{$(0,0)$ Configuration:} Uniform alignment across all molecules.
    \item \textbf{$(0,\pi)$ Configuration:} Alternating tautomers along the $x$-axis ($\Delta E = 5.45$~meV/mol).
    \item \textbf{$(\pi,0)$ Configuration:} Alternating tautomers along the $y$-axis ($\Delta E = 6.34$~meV/mol).
    \item \textbf{$(\pi,\pi)$ Configuration:} Checkerboard alignment ($\Delta E = 9.81$~meV/mol).
\end{itemize}

To model the macroscopic thermal ensemble at liquid-nitrogen temperature ($T = 77$~K, where $k_B T \approx 6.63$~meV), single-molecule Boltzmann weighting factors were computed as:
\begin{equation}
w_i = \frac{\exp(-\Delta E_{i,\text{mol}} / k_B T)}{\sum_{j} \exp(-\Delta E_{j,\text{mol}} / k_B T)}
\end{equation}
where $\Delta E_{i,\text{mol}} = (E_i^{\text{cell}} - E_{(0,0)}^{\text{cell}})/4$ represents the relative total energy per molecule relative to the $(0,0)$ ground state.

\begin{table}[h!]
\centering
\caption{\textbf{Relative energies and $77$~K Boltzmann populations for H$_2$Nc tautomer configurations.}}
\label{tab:tautomers}
\begin{tabular}{lcc}
\hline\hline
Configuration & $\Delta E_{\text{mol}}$ (meV/mol) & Boltzmann Weight ($77$~K) \\
\hline
$(0,0)$ & 0.00 & $48.71\%$ \\
$(0,\pi)$ & 5.45 & $21.44\%$ \\
$(\pi,0)$ & 6.34 & $18.74\%$ \\
$(\pi,\pi)$ & 9.81 & $11.11\%$ \\
\hline\hline
\end{tabular}
\end{table}

At $77$~K, thermal excitation readily populates all four configurations, with the $(0,0)$ ground state comprising nearly half ($48.71\%$) of the thermal ensemble. Because the energy differences between configurations are comparable to $k_B T$, the high-temperature limit approaches a spatially and temporally disordered ensemble that washes out microscopic $C_2$ anisotropy into an effective macroscopic $C_4$ symmetry.

Two limitations of this treatment should be noted. First, Boltzmann weighting over four ordered supercell configurations does not capture a genuinely disordered ensemble; with $\Delta E \sim k_B T$ the physical high-temperature limit approaches a random arrangement of tautomer orientations, and the weights in Table~\ref{tab:tautomers} should be read as bounding the magnitude of the averaging effect rather than as an exact ensemble average. Second, whether the tautomer distribution is thermally equilibrated on the timescale of a photoemission measurement is a kinetic question, governed by the hydrogen-transfer barrier and by tunnelling, rather than a thermodynamic one determined by the static energy differences computed here.~\cite{NC_switch,Naph_switch2}

Importantly, because the frontier HOMO orbital density resides primarily away from the central hydrogens, the HOMO dispersion and bandwidth ($3\text{--}4$~meV) are decoupled from the specific tautomeric arrangement. Consequently, static $(0,0)$ ground-state calculations accurately capture the active-space electronic structure of the thermal ensemble measured in experiment. We note that the residual in-plane anisotropy of the HOMO band is in any case small, $t_x/t_y = 0.92$; the LUMO-derived band carries a substantially larger anisotropy, $t_x/t_y = 1.59$, and is therefore the relevant state wherever tautomer ordering is discussed as a route to spatial anisotropy in the lattice model.

\begin{figure}[t!]
\centering
\includegraphics[width=0.7\textwidth]{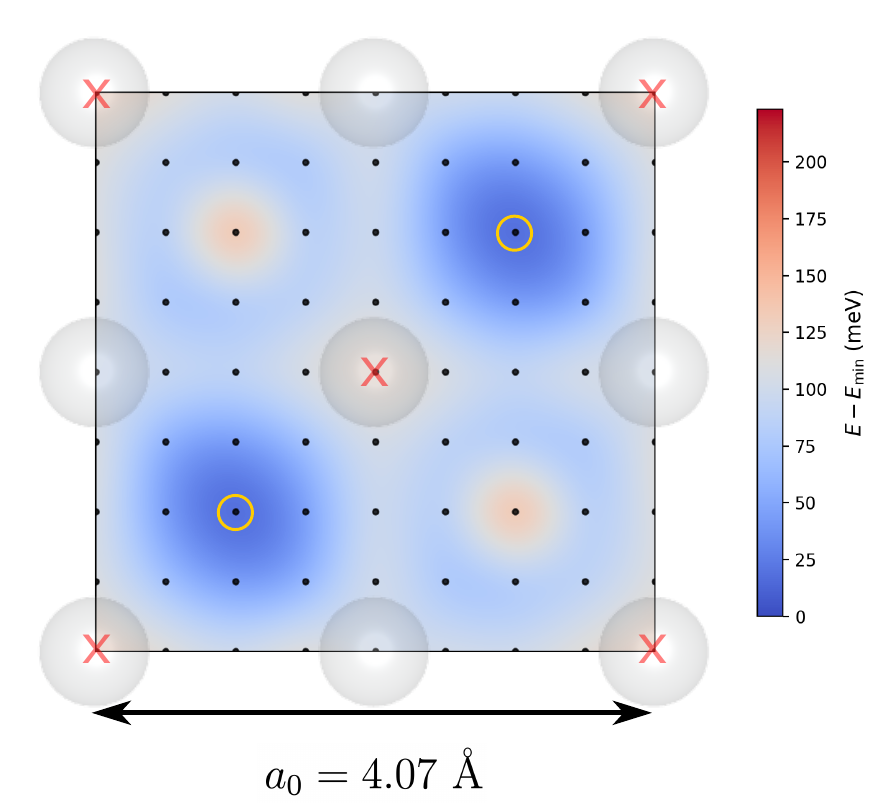}
\caption{\textbf{Potential energy surface (PES) for relative lateral displacements of the H$_2$Nc monolayer on Ag(100).} Relative total energy $E - E_{\text{min}}$ (meV per unit cell) plotted across a $0.125\,a_0$ relative translation grid. Black dots denote computed grid points sampled at a spatial resolution of $a_0/8 \approx 0.51$~\AA, relative to the bulk Ag lattice constant of $a_0=4.07$~\AA. Transparent grey spheres represent the surface Ag atoms: red crosses indicate top-layer Ag atoms, while unmarked spheres correspond to second-layer Ag atoms of the fcc(100) termination. Yellow circles highlight the symmetry-equivalent global potential energy minima corresponding to the ground-state adsorption registry.}
\label{fig:translations}
\end{figure}

The lateral registry of the monolayer on the substrate was determined by a rigid translation scan. Starting from the fully relaxed free-standing film placed above the Ag(100) slab, the molecular layer was translated over the two-dimensional space of relative in-plane displacements on a grid of spacing $0.125\,a_0 \approx 0.51$~\AA. At each translation, the $z$-coordinates of the atoms in the molecular layer were relaxed to a force convergence of 1~meV/\AA, using $\Gamma$-only $k$-point sampling. The resulting potential energy surface is shown in Fig.~\ref{fig:translations}. It spans approximately 220~meV per unit cell between the highest- and lowest-energy placements, so the registry is not a negligible degree of freedom, and it varies smoothly on the scale of the sampling grid, indicating that the located minimum is robust to finer sampling. The refined grid identifies two symmetry-equivalent global minima (yellow circles in Fig.~\ref{fig:translations}) lying $\approx 97$~meV per unit cell below the candidate registry obtained from a coarser $0.25\,a_0$ scan. All substrate-supported results reported here were computed on this refined registry. The minimum preserves the symmetric placement of the molecule on the fourfold Ag(100) surface net, so the only symmetry reduction in the adsorbed system is the intrinsic $C_2$ of the free H$_2$Nc macrocycle.

\subsection{Functional Dependence of Band Dispersion and Hopping Parameters}
\label{sec:hse_comparison}

\begin{table}[h!]
\centering
\caption{Comparison of fundamental gap $E_g$ and frontier band tight-binding hoppings $t_x, t_y$ calculated using PBE and HSE06 functionals for the free-standing H$_2$Nc monolayer.}
\label{tab:pbe_vs_hse}
\begin{tabular}{lccc}
\toprule
Parameter & PBE & HSE06 & Relative Change (\%) \\
\midrule
Band Gap $E_g$ (eV) & 1.08 & 1.37 & $+26.8\%$ \\
HOMO $t_x$ (meV)    & $-0.435$ & $-0.421$ & $-3.2\%$ \\
HOMO $t_y$ (meV)    & $-0.475$ & $-0.415$ & $-12.6\%$ \\
LUMO $t_x$ (meV)    & $-0.526$ & $-0.627$ & $+19.2\%$ \\
LUMO $t_y$ (meV)    & $-0.330$ & $-0.408$ & $+23.6\%$ \\
\bottomrule
\end{tabular}
\end{table}

To evaluate how exact exchange affects the band dispersions, band gap, and extracted tight-binding parameters of the free-standing H$_2$Nc monolayer, HSE06 calculations were performed self-consistently at the PBE-relaxed geometry, initialized from converged PBE orbitals, as described in Methods. Table~\ref{tab:pbe_vs_hse} compares the fundamental band gap and tight-binding hopping parameters extracted for the frontier HOMO and LUMO bands between PBE and HSE06. The HSE06 fundamental gap opens to 1.37~eV (compared to 1.08~eV in PBE), consistent with the well-known tendency of DFT-PBE to underestimate quasiparticle gaps in organic molecular solids.~\cite{Gap_renorm1,Gap_renorm2}

The appropriate experimental comparator for a calculated fundamental gap is a single-particle (transport) gap rather than an optical absorption gap, since optical gaps are reduced relative to the fundamental gap by the exciton binding energy, $\approx 0.3$--$0.5$~eV for this class of molecular film.~\cite{Binding_Energy} Two such comparators are available for H$_2$Nc. The first is the $\approx 1$~eV HOMO--LUMO separation resolved by ARPES for H$_2$Nc/Ag(100) in this work [Fig.~\ref{fig:Ag100}(a)], for which the PBE calculation on the same adsorbed system reproduces the measured value to within $\approx 0.1$~eV. The second is the $1.80\pm0.04$~eV single-particle gap from orbital-mediated tunneling spectroscopy on H$_2$Nc/Au(111),~\cite{Naph2} which is instead compared to our free-standing calculation and which both functionals underestimate, PBE (1.08~eV) more severely than HSE06 (1.37~eV). Because that measurement is made on a metal surface, it is itself reduced by image-charge screening that the free-standing calculation omits, and is therefore a lower bound on the free-standing gap. We attribute this difference to the residual self-interaction error of PBE, which is largest for the well-localized, molecular-like frontier orbitals of the free-standing film and is expected to be substantially reduced for the delocalized, hybridized HOMO-derived states of the metallic Ag(100)-adsorbed system discussed in the main text; we have not repeated the HSE06 calculation for the adsorbed system to verify this directly. We therefore report PBE-derived parameters throughout, consistent with its closer agreement with the same-interface Ag(100) measurement.

Non-local exchange modifies the kinetic dispersion of the frontier orbitals, increasing $|t|$ by 19\% ($t_x$) and 24\% ($t_y$) for the LUMO band, while the HOMO hoppings decrease by 3\% ($t_x$) and 13\% ($t_y$). The LUMO anisotropy is preserved $(t_x/t_y=1.59\rightarrow1.54)$, while the HOMO remains essentially isotropic in both functionals $(0.92\rightarrow1.01)$.

Because the on-site interaction $U$ is dominated by the localized intramolecular Wannier density, which remains nearly identical for the HOMO orbital between PBE and HSE06, and the hopping magnitude $t$ shifts by no more than $\sim 25\%$, the effective correlation parameter $U/t$ shifts by no more than this same $\sim$25\% and is unchanged at the order-of-magnitude level. This confirms that the system resides firmly within the strongly correlated regime regardless of functional choice.

Table~\ref{tab:tb_pbe_vs_hse_all} provides the complete set of fitted tight-binding parameters ($\varepsilon_n, t_{x,n}, t_{y,n}$) for all 16 bands near the Fermi level, comparing PBE and HSE06 side-by-side. The orbital-dependent energy shifts introduced by HSE06 reorder closely spaced levels relative to PBE. Bands in Table~\ref{tab:tb_pbe_vs_hse_all} are indexed strictly by ascending energy within each functional, so index $n$ does not correspond to the same orbital in both columns; for example, the orbital character of PBE band 1 maps onto HSE06 band 2. Row-by-row comparison of the non-frontier bands in Table~\ref{tab:tb_pbe_vs_hse_all} is therefore not meaningful, and the large apparent changes in $t_{x,n}$ and $t_{y,n}$ for several of these rows, including sign reversals, reflect this reindexing rather than a functional dependence of the hoppings themselves. The frontier HOMO and LUMO bands are unambiguously identified by their gap position and orbital character in both functionals and are the only rows for which we draw a quantitative comparison. For ease of comparison, the HSE06 spectrum has been rigidly shifted so that its HOMO eigenvalue coincides with the PBE HOMO. The shift is a bookkeeping choice and carries no physical content: only relative level spacings, and not absolute eigenvalues, should be compared between the two functionals.

\begin{table}[htbp]
\centering
\small
\setlength{\tabcolsep}{3.5pt}
\caption{Side-by-side comparison of tight-binding parameters calculated using PBE and HSE06 functionals for the 16 bands nearest to the Fermi level in the free-standing $\mathrm{H}_2\mathrm{Nc}$ monolayer: on-site energies $\varepsilon_n$, nearest-neighbor hoppings $t_{x,n}$ and $t_{y,n}$ along the $x$ and $y$ directions, and fit quality $R^2$. Bands are indexed by ascending energy within each functional; see text regarding index matching between functionals. For ease of comparison, the HSE06 spectrum has been rigidly shifted so that its HOMO eigenvalue coincides with the PBE HOMO, which does not change the physical interpretation of the results.}
\label{tab:tb_pbe_vs_hse_all}
\begin{tabular}{c *{8}{r}}
\toprule
 & \multicolumn{2}{c}{$\varepsilon_n$ (eV)} & \multicolumn{2}{c}{$t_{x,n}$ (meV)} & \multicolumn{2}{c}{$t_{y,n}$ (meV)} & \multicolumn{2}{c}{$R^2$} \\
\cmidrule(lr){2-3} \cmidrule(lr){4-5} \cmidrule(lr){6-7} \cmidrule(lr){8-9}
Band $n$ & {PBE} & {HSE06} & {PBE} & {HSE06} & {PBE} & {HSE06} & {PBE} & {HSE06} \\
\midrule
1  & -2.318 & -2.895 &  0.446 & -0.018 &  0.451 &  0.295 & 0.987 & 0.997 \\
2  & -2.211 & -2.739 & -1.172 &  0.464 & -1.285 &  0.412 & 0.999 & 0.986 \\
3  & -2.091 & -2.621 & -0.602 & -1.365 &  0.037 & -1.241 & 0.999 & 0.999 \\
4  & -2.080 & -2.517 & -0.151 &  0.041 & -0.168 & -0.545 & 0.999 & 0.998 \\
5  & -2.046 & -2.479 & -0.385 & -0.367 & -0.010 & -0.034 & 0.995 & 0.992 \\
6  & -1.722 & -2.105 & -0.718 & -0.809 &  0.260 &  0.229 & 0.987 & 0.988 \\
7  & -1.631 & -2.006 &  2.197 &  2.297 &  2.167 &  2.207 & 0.998 & 0.996 \\
8  & -1.614 & -1.992 & -0.178 & -0.179 & -0.773 & -0.819 & 0.971 & 0.951 \\
9 (HOMO) & -0.552 & -0.552 & -0.435 & -0.421 & -0.475 & -0.415 & 0.997 & 0.997 \\
10 (LUMO) &  0.531 &  0.816 & -0.526 & -0.627 & -0.330 & -0.408 & 0.996 & 0.997 \\
11 &  0.627 &  0.934 & -0.384 & -0.462 & -0.981 & -1.132 & 1.000 & 1.000 \\
12 &  1.368 &  1.863 & -0.047 & -0.138 & -0.134 & -0.176 & 0.947 & 0.966 \\
13 &  1.569 &  2.091 & -2.814 & -3.445 &  0.452 &  0.641 & 0.998 & 0.998 \\
14 &  1.628 &  2.151 &  0.413 &  0.531 &  0.547 &  0.624 & 0.989 & 0.982 \\
15 &  1.710 &  2.241 &  0.269 &  0.402 & -2.293 & -2.833 & 0.998 & 0.998 \\
16 &  1.988 &  2.510 &  4.026 &  5.046 &  3.699 &  4.671 & 0.999 & 0.998 \\
\bottomrule
\end{tabular}
\end{table}

\subsection{Adequacy of the nearest-neighbor model}
\label{sec:nnn_model}
Across the 16-band window, two bands fall below $R^2 = 0.98$ under the nearest-neighbor-only fit in either functional: band 8 ($R^2 = 0.971$ in PBE, $0.951$ in HSE06) and band 12 ($R^2 = 0.947$ in PBE, $0.966$ in HSE06). All remaining bands, including both frontier orbitals, are fitted at $R^2 \geq 0.98$; the HOMO and LUMO reach $R^2 = 0.997$ and $0.996$ respectively.

To confirm that the reduced fit quality for these two bands reflects a genuine next-nearest-neighbor (diagonal) hopping rather than a breakdown of the tight-binding description, we refit them with the extended anisotropic model
\begin{equation}
E_n(\mathbf{k}) = \varepsilon_n + 2t_{x,n}\cos(k_x a_x) + 2t_{y,n}\cos(k_y a_y) + 4t'_n \cos(k_x a_x)\cos(k_y a_y),
\label{eq:nnn_model}
\end{equation}
where $t'_n$ is the hopping along the $(\pm 1, \pm 1)$ lattice diagonal. The results are given in Table~\ref{tab:nnn_fits}. Including a single additional parameter recovers $R^2 = 0.998$ for band 12, confirming that its residual under the nearest-neighbor-only fit is attributable to next-nearest-neighbor (diagonal) hopping. Band 8 is essentially unchanged by the same extension ($R^2 = 0.971 \to 0.972$), indicating that its modest residual is not captured by diagonal next-nearest-neighbor hopping and more plausibly reflects a weak coupling to more distant neighbors. 

We nonetheless retain the minimal nearest-neighbor model throughout Table~\ref{tab:tb_pbe_vs_hse_all}, in order to preserve a uniform basis of comparison across all 16 bands and between the two functionals. Because neither band 8 nor band 12 lies in the frontier active space near $E_F$, and neither enters the $U/t$ analysis, this choice does not affect any conclusion of the main text. 

\begin{table}[h!]
\centering
\caption{Extended nearest- and next-nearest-neighbor fits (Eqn.~\ref{eq:nnn_model}) for the two bands whose nearest-neighbor-only fit falls below $R^2 = 0.98$. PBE values.}
\label{tab:nnn_fits}
\begin{tabular}{crrrrr}
\toprule
Band $n$ & $t_{x,n}$ (meV) & $t_{y,n}$ (meV) & $t'_n$ (meV) & $R^2$ (NN only) & $R^2$ (NN + NNN) \\
\midrule
8  & -0.178 & -0.773 & -0.014 & 0.971 & 0.972 \\
12 & -0.047 & -0.134 & -0.020 & 0.947 & 0.998 \\
\bottomrule
\end{tabular}
\end{table}

\subsection{Bloch Orbital Parity}
\label{sec:bloch_parity}

One may notice that the parity (symmetric/antisymmetric) of the Bloch orbitals flips for some of the orbitals in Fig.~\ref{fig:bands_and_wfns}(f) and (g) of the main text between the film and the isolated molecule. Because the Bloch states at $\Gamma$ in the film are formed by periodic combinations of molecular orbitals, near-degenerate levels can recombine into bonding/antibonding mixtures which differ in local parity relative to the isolated molecule. This apparent parity flip is a numerical consequence of the arbitrary orthonormal basis returned by diagonalization in a near-degenerate subspace; physical observables are unaffected. 

\subsection{Coulomb Interactions Calculation Method}
\label{sec:coulomb_method}

Coulomb integrals for the HOMO maximally localized Wannier function $w(\mathbf{r})$ were evaluated using a fast Fourier-transform (FFT) scheme used to generate the full two-dimensional interaction map $U(\mathbf{R})$. In the FFT approach, the in-plane density $\rho_2(x,y)=\int dz\,|w(x,y,z)|^2$ was constructed on the real-space grid exported from the Wannier/XSF file, centered to sub-pixel accuracy by applying an FFT phase shift to the complex amplitude, truncated to a box enclosing the orbital, and then zero-padded by integer pad factors to increase the in-plane supercell size $L$; the smallest Fourier wavevector was therefore $q_{\textrm{min}}=2\pi/L$. The two-dimensional Fourier transform $\rho_{\mathbf{q}}$ was computed using NumPy's \texttt{fft2}, the screened interaction kernel $V(\mathbf{q})$ in Eqn.~(\ref{eq:kernel}) of the main text was evaluated on the discrete $\mathbf{q}$ mesh, and $U(\mathbf{R})$ was obtained from the inverse FFT of $V(\mathbf{q})|\rho_{\mathbf{q}}|^2$. Analytic small-$\mathbf{q}$ handling was used to avoid the singularity at $|\mathbf{q}|=0$. For the free-standing limit ($\xi\rightarrow\infty$), calculations were performed for a sequence of pad factors and the finite-$L$ results were extrapolated to $L\rightarrow\infty$ using a fit of the form $U(L)=U_\infty+a/L$. Finite-$\xi$ results reported in the main text were computed using a pad factor of 8. 

The kernel of Eqn.~(\ref{eq:kernel}) of the main text is written for an isotropic background, but layered substrates such as hBN are uniaxial. For a monolayer at $z=0$ between grounded gates at $z=\pm\xi$ in a uniform uniaxial medium, Poisson's equation $\varepsilon_{\parallel}(\partial_x^2+\partial_y^2)\phi +\varepsilon_{\perp}\partial_z^2\phi = -\rho/\varepsilon_0$ gives, in-plane Fourier space, $\phi \propto e^{\mp\kappa z}$ with $\kappa = q\sqrt{\varepsilon_{\parallel}/\varepsilon_{\perp}}$: the anisotropy enters only as a rescaled decay constant.\cite{LandauLifshitzECM} Imposing $\phi(q,\pm\xi)=0$ and the displacement-field jump at the charge sheet yields
\[
V(q) = \frac{e^{2}}
{2\varepsilon_{0}\sqrt{\varepsilon_{\parallel}\varepsilon_{\perp}}\,|q|}
\tanh\!\left(|q|\,\xi\sqrt{\varepsilon_{\parallel}/\varepsilon_{\perp}}\right),
\]
identical to Eqn.~(\ref{eq:kernel}) under $\varepsilon \rightarrow \tilde{\varepsilon} = \sqrt{\varepsilon_{\parallel}\varepsilon_{\perp}}$ and $\xi \rightarrow \xi\sqrt{\varepsilon_{\parallel}/\varepsilon_{\perp}}$, and reducing to it when $\varepsilon_{\parallel}=\varepsilon_{\perp}$. The reduction is exact within the idealization Eqn.~(\ref{eq:kernel}) already assumes (homogeneous medium, zero-thickness sheet, ideal gates); only in-plane biaxial anisotropy ($\varepsilon_{xx}\neq\varepsilon_{yy}$) falls outside it. For hBN ($\varepsilon_{\parallel}\approx6.9$, $\varepsilon_{\perp}\approx3.8$)\cite{Laturia2018} this gives $\tilde{\varepsilon}\approx5.1$ and $\tilde{\xi}\approx1.35\,\xi$, so anisotropy relocates a device along the curves of Fig.~\ref{fig:Hubbard_U}(d) rather than reshaping them. Anisotropy of the surrounding environment is distinct from the in-plane polarizability of the film itself, which is the same channel as the intramolecular screening making $U_{\mathrm{bare}}$ an upper bound; the two should not be applied together.

%\newpage

\begin{figure}
    \includegraphics[width=.99\textwidth]{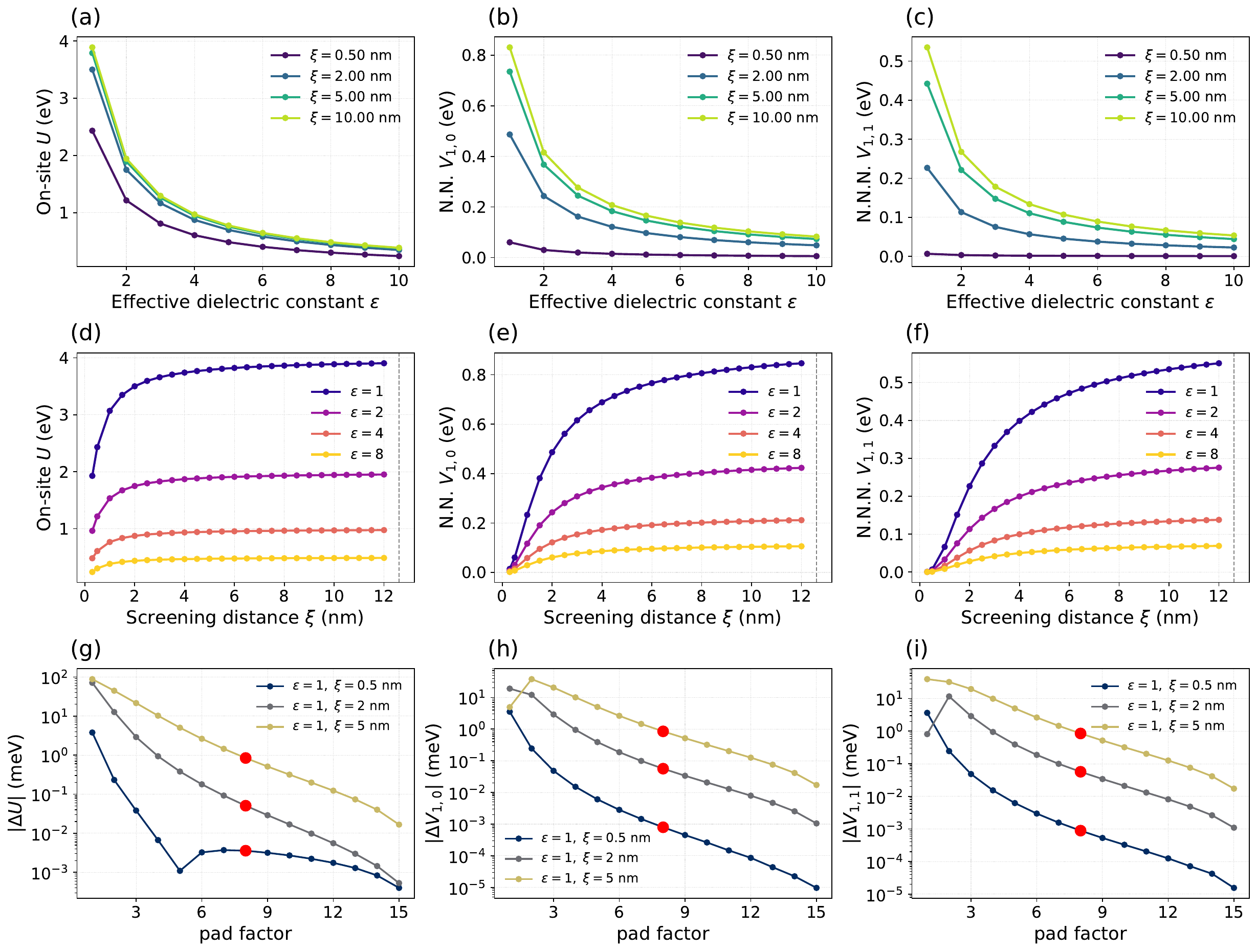}  
    \caption{Coulomb matrix elements $U$, $V_{1,0}$, and $V_{1,1}$ computed from the H$_2$Nc HOMO Wannier orbital as functions of $\varepsilon$ and $\xi$. The bottom row shows the absolute convergence error relative to the largest padding factor tested, which was 16. The red markers show errors for the pad factor of 8, which was used for finite-$\xi$ results presented in this paper.}
    \label{fig:Hubbard_U_Extra}
\end{figure}

\subsection{Determination of Effective Screened Hopping $t_{\text{screened}}$ on Ag(100)}
\label{sec:t_screened}

For the adsorbed $\text{H}_2\text{Nc}/\text{Ag}(100)$ monolayer, strong molecule-substrate hybridization causes the molecular HOMO character to spread across several near-degenerate bands simultaneously at a given wavevector $\mathbf{k}$, rather than residing in a single, uniquely identifiable eigenvalue. This precludes both a direct single-band tight-binding fit to individual dispersion eigenvalues $E_n(\mathbf{k})$ and a fixed-energy-window average of the molecular weight, the latter of which we found to be numerically unstable: restricting to a narrow, percentile-defined energy window produced total molecular weight $W_{\text{tot}}(\mathbf{k})$ varying by 30--150\% across the sampled zone, with some $\mathbf{k}$-points losing essentially all weight from the window entirely.

We instead treat the problem as one of extracting a dispersion from a continuous spectral intensity, in direct analogy with how a photoemission band is identified from an energy distribution curve. We define a molecular-character-weighted, Gaussian-broadened spectral function
\begin{equation}
A(\mathbf{k}, E) = \sum_{n} w_n^{\mathrm{mol}}(\mathbf{k}) \, \frac{1}{\sigma\sqrt{2\pi}} \exp\!\left[-\frac{(E - E_n(\mathbf{k}))^2}{2\sigma^2}\right], 
\label{eq:spectral_function}
\end{equation}
where $w_n^{\mathrm{mol}}(\mathbf{k})$ is the total molecular (C, N, H) projection weight of band $n$ at $\mathbf{k}$, and $\sigma$ is a post-processing broadening applied when constructing $A(\mathbf{k}, E)$ from a single, fixed set of already-converged DFT eigenvalues $E_n(\mathbf{k})$ and projection weights $w_n^{\mathrm{mol}}(\mathbf{k})$. We emphasize that $\sigma$ is unrelated to the electronic smearing used during the self-consistent DFT calculation (Methods): the underlying band structure is computed once, and $\sigma$ only controls how that fixed result is subsequently processed into a continuous intensity for peak-tracking. At each $\mathbf{k}$ we identify the dominant peak of $A(\mathbf{k}, E)$ using a prominence-based criterion (retaining peaks with prominence $\geq 20\%$ of the local maximum, to distinguish genuine features from fine sub-structure arising from near-degenerate bands), and track its energy, $E_{\text{peak}}(\mathbf{k})$, across the Brillouin zone. Because $A(\mathbf{k}, E)$ is a continuous field rather than a discrete band index, this construction is insensitive to avoided crossings: near a crossing, spectral weight redistributes continuously between nearby features rather than requiring a discrete choice between them, in contrast to both a fixed-window average and a nearest-energy continuity-tracking scheme, both of which we found to give unstable results for this strongly hybridized system.

The active-space bandwidth is then
\begin{equation}
\Delta E_{\text{disp}} = \max_{\mathbf{k}} E_{\text{peak}}(\mathbf{k}) - \min_{\mathbf{k}} E_{\text{peak}}(\mathbf{k}),
\label{eq:peak_bandwidth}
\end{equation}
and the effective screened hopping follows from the isotropic nearest-neighbor relation $W = 8t$ for the square lattice,
\begin{equation}
t_{\text{screened}} = \frac{\Delta E_{\text{disp}}}{8}.
\label{eq:t_screened_final}
\end{equation}

\begin{figure}
    \includegraphics[width=.99\textwidth]{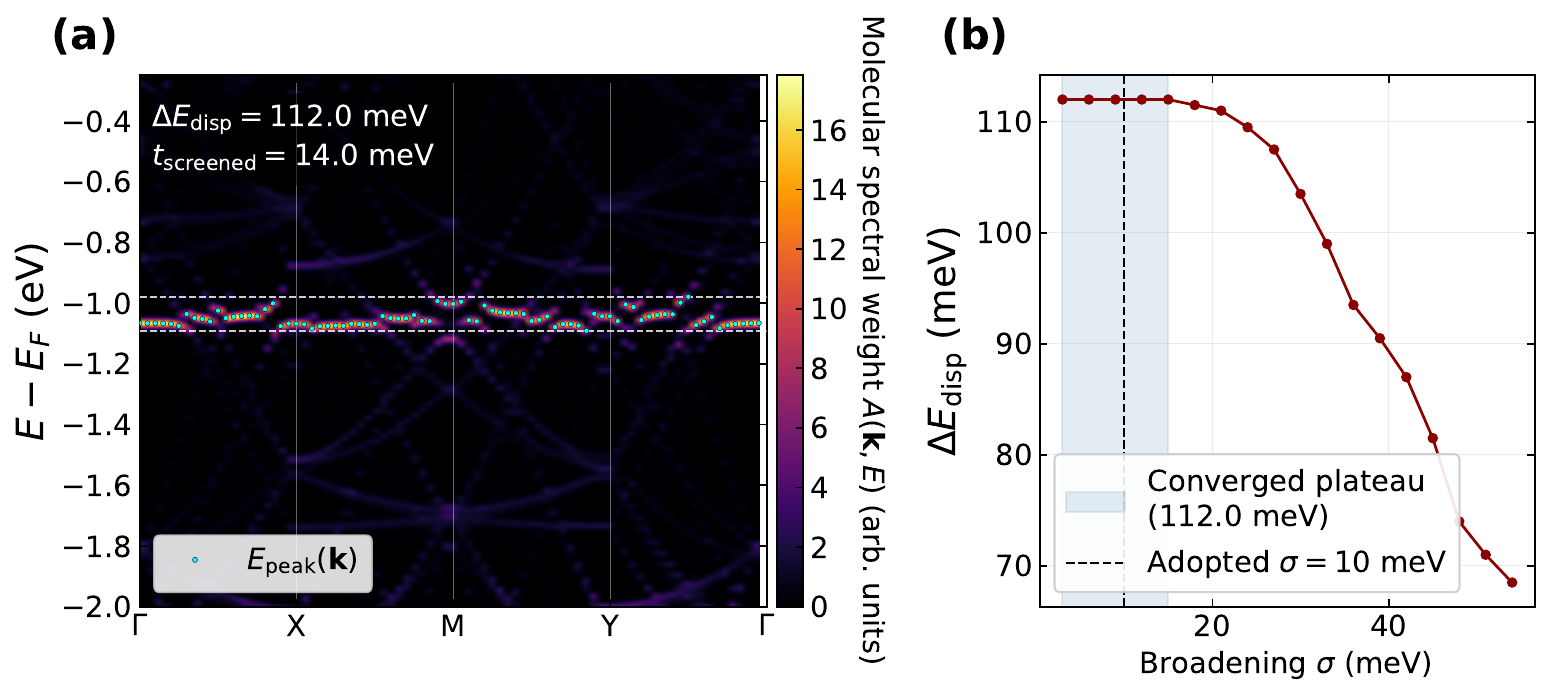}  
    \caption{\textbf{Extraction of the screened hopping $t_{\text{screened}}$ from a molecular-projected spectral function.} (a) Molecular-character-weighted spectral function $A(\mathbf{k}, E) = \sum_n w_n^{\mathrm{mol}}(\mathbf{k})\, G(E - E_n(\mathbf{k}); \sigma)$ for the H$_2$Nc/Ag(100) HOMO manifold, evaluated at the adopted broadening $\sigma = 10$~meV. The dominant peak at each $\mathbf{k}$, $E_{\text{peak}}(\mathbf{k})$ (cyan), traces a single feature across the Brillouin zone despite the underlying molecular weight being fragmented across several near-degenerate bands at fixed $\mathbf{k}$. Dashed white lines mark the maximum and minimum of the trace, defining $\Delta E_{\text{disp}} = 112.0$~meV and, via $W = 8t$, $t_{\text{screened}} = 14.0$~meV. (b) Convergence of the extracted bandwidth $\Delta E_{\text{disp}}$ with respect to the broadening $\sigma$. The shaded region marks the range over which the extracted value is converged to within 1~meV ($\sigma = 3$--$15$~meV); for $\sigma \gtrsim 20$~meV the broadening kernel begins to merge dispersion across different $\mathbf{k}$, progressively narrowing the apparent bandwidth rather than resolving additional spurious structure. The value reported in the main text is taken from this plateau, at $\sigma = 10$~meV (dashed line).}
\label{fig:tscreened_spectral}
\end{figure}

Figure~\ref{fig:tscreened_spectral}(a) shows $A(\mathbf{k}, E)$ evaluated at $\sigma = 10$~meV together with the tracked peak $E_{\text{peak}} (\mathbf{k})$, which follows a single, continuous, periodically modulated feature across the path. Figure~\ref{fig:tscreened_spectral}(b) shows the convergence of $\Delta E_{\text{disp}}$ with respect to $\sigma$: the extracted bandwidth is stable to better than 1~meV for $\sigma = 3$--$15$~meV ($\Delta E_{\text{disp}} = 112.0$~meV), and decreases smoothly for larger broadening as $\sigma$ approaches the intrinsic energy scale of the dispersion itself, causing the kernel to merge spectral weight across different $\mathbf{k}$ rather than resolving genuine sub-structure. We report the plateau value, $\sigma = 10$~meV, giving
\begin{equation}
t_{\text{screened}} = \frac{112.0~\text{meV}}{8} = 14.0~\text{meV}.
\end{equation}

This construction isolates the lateral momentum dispersion of the adsorbed monolayer from the non-dispersive lifetime broadening imparted by the metal, and provides the baseline effective hopping scale used in evaluating the screened $U/t$ ratio on Ag(100).

\subsection{Interfacial Charge Redistribution and Charge Transfer Analysis}
\label{sec:charge_transfer}

To quantitatively characterize the electronic interaction and substrate-to-adsorbate charge transfer at the interface, we computed the plane-averaged charge density difference along the surface normal ($z$-axis),~\cite{Peisert2002,Khomyakov2009} defined as:
\begin{equation}
\Delta\bar{\rho}(z) = \bar{\rho}_{\text{H}_2\text{Nc/Ag}}(z) - \bar{\rho}_{\text{Ag}}(z) - \bar{\rho}_{\text{H}_2\text{Nc}}(z),
\end{equation}
where $\bar{\rho}_{\text{H}_2\text{Nc/Ag}}(z)$, $\bar{\rho}_{\text{Ag}}(z)$, and $\bar{\rho}_{\text{H}_2\text{Nc}}(z)$ represent the $xy$-plane averaged total electron densities of the full adsorbed system, the clean isolated Ag(100) substrate, and the freestanding $\mathrm{H_2Nc}$ monolayer, respectively. For the reference systems, atomic positions were held fixed at their fully relaxed interfacial geometry within identical supercell dimensions.

To evaluate the spatial redistribution of charge, the cumulative charge transfer $\Delta Q(z)$ integrated along the surface normal is given by:
\begin{equation}
\Delta Q(z) = \int_{z_0}^{z} \Delta\bar{\rho}(z') \, dz',
\end{equation}
where $z_0$ is chosen deep within the vacuum region below the bottom of the Ag substrate slab. 

\begin{figure}[htbp]
    \centering
    \includegraphics[width=0.85\linewidth]{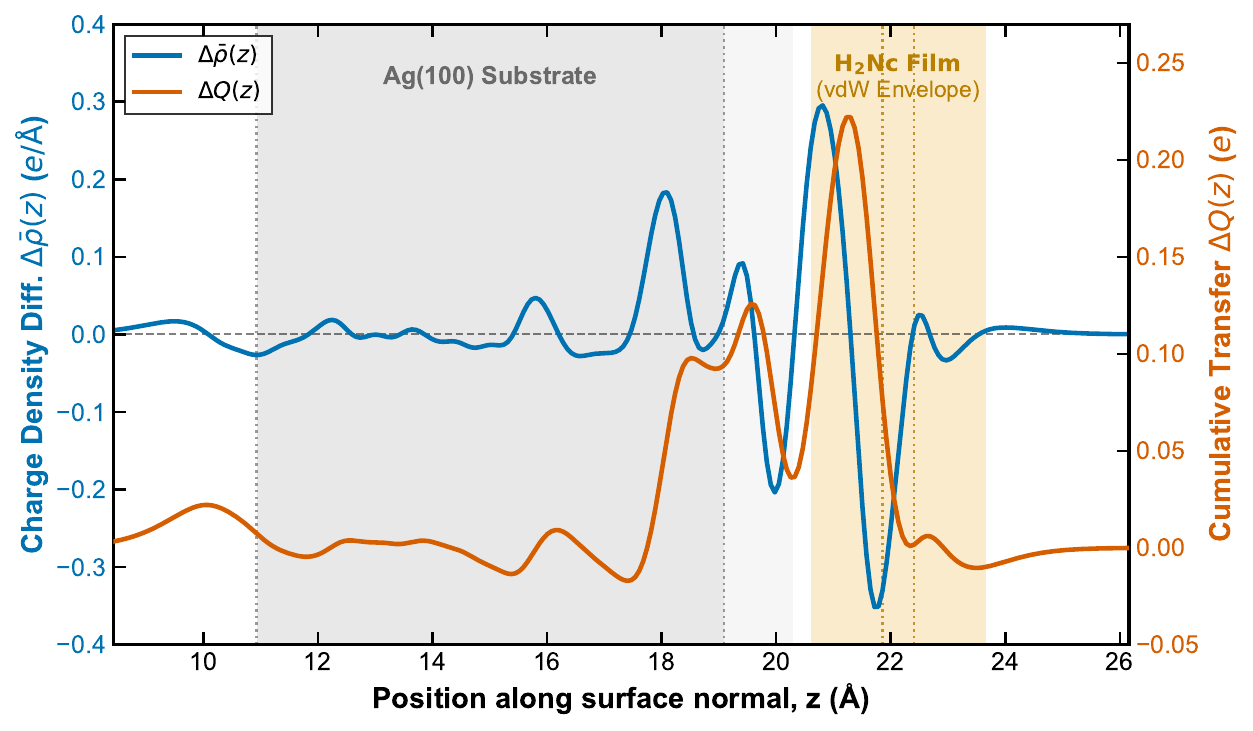}
    \caption{\textbf{Quantitative interfacial charge redistribution along the surface normal.} 
    Plane-averaged charge density difference $\Delta\bar{\rho}(z)$ (blue curve, left axis) and corresponding cumulative charge transfer $\Delta Q(z)$ (vermilion curve, right axis) plotted along the surface normal $z$. Solid grey region denotes the bulk atomic layers of the Ag(100) substrate slab, bounded by vertical dotted lines marking nuclear planes ($10.93$~\AA\ to $19.09$~\AA). The hatched grey extension ($19.09$~\AA\ to $20.29$~\AA) highlights the spilled-out surface electron density tail of the metal. The yellow shaded region highlights the van der Waals envelope ($\approx 1.25$~\AA\ extension) surrounding the nuclear framework of the $\mathrm{H_2Nc}$ monolayer ($21.86$~\AA\ to $22.41$~\AA, vertical dotted lines). The distinct positive peak of $\Delta\bar{\rho}(z) \approx +0.30$~$e$/\AA\ at $z \approx 20.8$~\AA\ corresponds to local accumulation in the lower $\pi$-lobes (LUMO back-donation), associated with a localized cumulative charge transfer peak of $\Delta Q(z) \approx +0.22\,e$. Strong intramolecular polarization produces a subsequent negative depletion region within the upper molecular framework.}
    \label{fig:charge_transfer}
\end{figure}

As shown in Fig.~\ref{fig:charge_transfer}, the charge redistribution exhibits a pronounced dipolar behavior at the interface. At $z \approx 20.8$~\AA, within the lower van der Waals envelope of the monolayer, $\Delta\bar{\rho}(z)$ reaches a sharp positive peak of $\approx +0.30\,e/\text{\AA}$, yielding a localized accumulation peak of $\Delta Q(z) \approx +0.22\,e$. This confirms significant interfacial charge redistribution into the lower lobes of the molecular lowest unoccupied molecular orbital (LUMO). Directly above this accumulation layer ($z \approx 21.8$~\AA), a localized charge depletion region ($\Delta\bar{\rho}(z) \approx -0.35\,e/\text{\AA}$) reflects intramolecular $\sigma$-polarization driven by the strong surface dipole field. 

While the 1D continuum integral $\Delta Q(z)$ captures this localized $0.22\,e$ accumulation at the immediate interface, the full 3D net molecular charge computed via DDEC6~\cite{DDEC6_1,DDEC6_2} iterative Stockholder partitioning yields a smaller net gain of $+0.10\,e$ per H$_2$Nc molecule. This difference arises because atomic volume partitioning schemes aggregate electron density over the entire 3D molecular basin, effectively summing the localized $0.22\,e$ LUMO back-donation alongside the opposing $\approx -0.12\,e$ intramolecular $\sigma$-depletion. Consequently, while DDEC6 accurately reports a near-neutral overall molecular charge, the 1D real-space density integration is essential to capture the magnitude of the local interfacial charge transfer driving the bonding.

\subsubsection{Dependence on the charge partitioning scheme}

Because the inferred charge transfer at a metal--organic interface is known to depend strongly on how the electron density is partitioned between the adsorbate and the substrate, we evaluated three independent measures on the same converged charge density and identical ground-state adsorption registry. These are collected in Table~\ref{tab:charge_partitions}.

\begin{table}[h!]
\centering
\caption{\textbf{Charge transfer at the H$_2$Nc/Ag(100) interface evaluated by three independent methods}, applied to the same converged charge density on the refined ground-state adsorption registry. All values are electrons transferred to the molecule; methods differ in how much of the interfacial charge redistribution each integration domain captures.}
\label{tab:charge_partitions}
\begin{tabular}{lcl}
\hline\hline 
Method & Charge ($e^-$/molecule) & Integration domain \\
\hline 
Bader (zero-flux topological)  & 0.82 & Full 3D atomic basin \\ DDEC6 (iterative Stockholder)  & 0.10 & Full 3D atomic basin, \\ & & \quad reference-constrained \\ $\Delta Q(z)$ interfacial peak & 0.22 & 1D planar integral, truncated \\ &      & \quad at the accumulation maximum \\
\hline\hline
\end{tabular}
\end{table}

The factor-of-eight spread between the Bader~\cite{Bader1,Bader2,Bader3,Bader4} and DDEC6~\cite{DDEC6_1,DDEC6_2} net charges illustrates how sensitively the inferred transfer depends on the partitioning convention. We regard the larger Bader value as an artifact of the scheme rather than a physical transfer: at a metal--organic interface, the zero-flux surface separating the molecule from the substrate encloses part of the diffuse metallic surface density spilling out of the top Ag layer, which is then assigned in its entirety to the molecular basin. This is visible directly in Fig.~\ref{fig:charge_transfer} as the spilled-out density tail in the hatched region between $19.09$~\AA\ and $20.29$~\AA. DDEC6 is substantially more robust against this boundary artifact because it constrains the partitioned atomic densities to be simultaneously consistent with the total density and with reference free-atom densities. The $\Delta Q(z)$ value captures only the LUMO back-donation peak and does not include the subsequent intramolecular $\sigma$-depletion ($\approx -0.12\,e^-$) that lies further from the interface within the same molecule.

We therefore base the physical interpretation in the main text on the DDEC6 net charge together with the partition-free $\Delta Q(z)$ integral, and report the Bader value for completeness and for comparison with the wider literature, where Bader charges are commonly quoted for this class of interface.

\subsection{Lattice models and requirements for quantum simulation}
\label{sec:simulation}

The parameter ranges reported in the main text map onto an extended square-lattice Hubbard model,
\begin{equation}
H = -\sum_{\langle ij \rangle, \sigma} t_{ij} \left(c_{i\sigma}^{\dagger} c_{j\sigma} + \mathrm{h.c.}\right) + U \sum_i n_{i\uparrow} n_{i\downarrow} + \frac{1}{2}\sum_{i \neq j} V_{ij}\, n_i n_j ,
\label{eq:si_extended_hubbard}
\end{equation}
in which $U$, $V_{1,0}$, and $V_{1,1}$ are set by the dielectric environment through Eqn.~\ref{eq:kernel} of the main text and $t$ by substrate hybridization. We define the filling as the number of electrons per molecular site, $n = N_e/N_{\mathrm{sites}}$; the HOMO-derived band is spinful, so $n \in [0,2]$, with $n = 1$ half filling and $n = 1/2$ quarter filling. This is the convention in which the square-lattice Mott boundary ($U_c/t \sim 8$--$12$)~\cite{Mott_Ratio} and the $t$--$J$ mapping are quoted.

\subsubsection*{Requirements}

Realizing the regimes summarized in the main text places four concrete demands on the device geometry.

\emph{Gating requires a decoupled geometry.} A bulk metal screens an applied gate field, so electrostatic filling control presupposes an insulating spacer between the H$_2$Nc layer and any metallic electrode, for which $\varepsilon$ and $\xi$ are the controlling parameters. At a site density of $3.8 \times 10^{13}$~cm$^{-2}$, sweeping $n$ across $[0,2]$ requires $\sim 10^{14}$~cm$^{-2}$ of induced carriers, beyond conventional SiO$_2$ back-gating and in the range of ionic-liquid or high-$\kappa$ gating.\cite{Ueno2008,Ye2012}

\emph{Interactions and hopping must be read at the same screening point.} The bare values $V_{1,0} = 0.930$~eV and $V_{1,1} = 0.634$~eV apply in the unscreened limit and are reduced by approximately the same factor as $U$ under screening, so the condition $V_{ij}/t \gg 1$ must be read from a single point in the $(\varepsilon,\xi)$ parameter space rather than by combining bare interactions with screened hopping.

\emph{The quoted exchange scale is an upper bound.} The scale $J = 4t_{\mathrm{screened}}^{2}/U_{\mathrm{screened}} \approx 3.9$~meV pairs $t_{\mathrm{screened}}$, computed for the H$_2$Nc/Ag(100) contact geometry, with $U$ at the most strongly screened point of Fig.~\ref{fig:Hubbard_U}(d). A fully decoupled film returns $t$ toward $t_{\mathrm{bare}} \approx 0.45$~meV and reduces $J$ accordingly, so this value bounds rather than predicts the achievable exchange.

\emph{Kondo screening and the absence of two-dimensional order.} On a bare noble-metal surface the hybridization width and associated Kondo scale compete with $J$, and local-moment screening rather than intersite exchange is expected to dominate.\cite{Zhao2005,Mugarza2011,Granet2020} The Heisenberg regime therefore also belongs to the decoupled geometry. Because the isotropic two-dimensional Heisenberg model supports no finite-temperature ordering transition,\cite{MerminWagner1966} the accessible observable is short-range antiferromagnetic correlation and its spin-fluctuation spectrum, with long-range order requiring residual anisotropy or interlayer coupling.

\subsection{Repeatability of the fitted dispersion of the H$_2$Nc/Au(111) HOMO band}
\label{sec:Au111_repeatability}

\begin{figure}[h]
    \centering
    \includegraphics[width=0.85\linewidth]{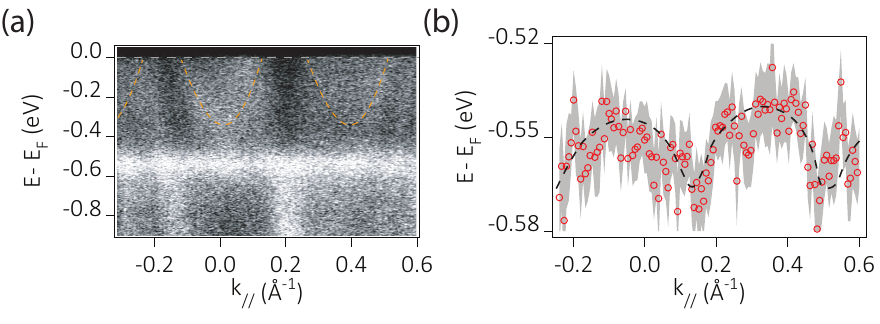}
    \caption{\textbf{Energy-momentum spectrum and the fitted dispersion of H$_2$Nc/Au(111) rotated away from normal emission.} 
    (a) The energy-momentum spectrum near $\Gamma$ but with the sample rotated for $5^{\circ}$. Yellow dashed lines denote replicas of the Au(111) Shockley surface state. (b) Energy dispersion of the HOMO band fitted from (a) using the same protocol as in the main text.}
    \label{fig:Au111_repeatability}
\end{figure}

Fig.~\ref{fig:Au111_repeatability} shows the ARPES spectrum and the corresponding fitted energy dispersion of the HOMO band near $\Gamma$ of H$_2$Nc/Au(111) but with the sample rotated away from normal photoemission for $5^{\circ}$. Compared to Fig.~\ref{fig:Au111}(b) and (d) in the main text, the overall shift of both the surface state replicas and the periodic energy dispersion to the left of the field of view demonstrates
that the features are not measurement artifacts and are repeatable.

\subsection{Fitting results of the HOMO band of H$_2$Nc/Ag(100)}
\label{sec:Ag100_fitting}
\begin{figure}[h]
    \centering
    \includegraphics[width=0.85\linewidth]{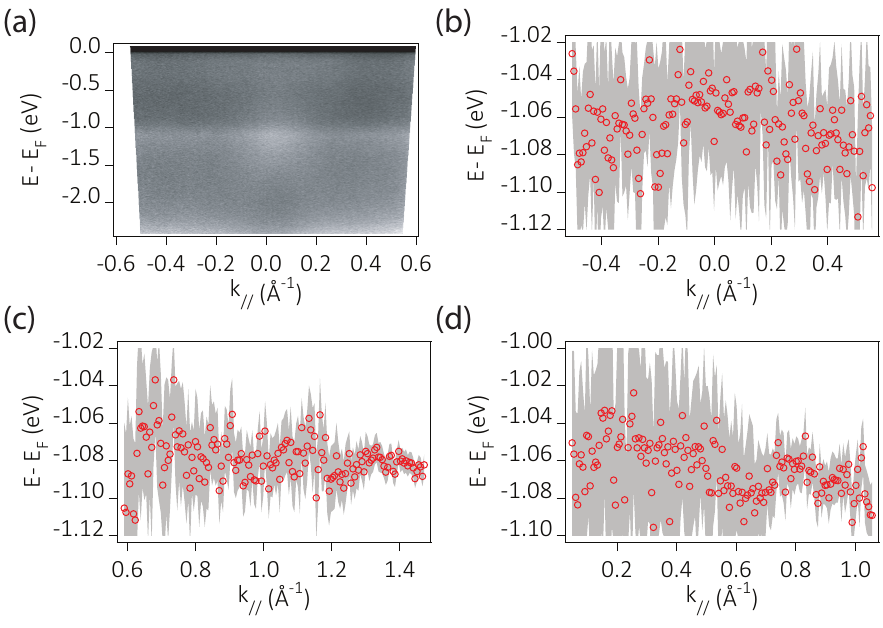}
    \caption{\textbf{Fitting results of the HOMO band of H$_2$Nc/Ag(100).} 
    (a) The energy-momentum spectrum near $\Gamma$ and (b) the fitting results of the HOMO energy positions in (a). (c) and (d) The fitting results of the HOMO energy positions in Fig.~\ref{fig:Ag100}(b) and (d).}
    \label{fig:Ag100_fitting}
\end{figure}

The single-band tight-binding fit shown in Fig.~\ref{fig:Ag100_fitting} is included for completeness but should be read qualitatively. The fitted dispersion is less well-defined than the H$_2$Nc/Au(111) case (SI Section~\ref{sec:Au111_repeatability}): the H$_2$Nc/Ag(100) HOMO peak carries a FWHM of 350~meV in the UPS spectrum (Fig.~\ref{fig:Ag100}(a) inset), roughly $3.5\times$ broader than on Au(111) (100~meV, Fig.~\ref{fig:Au111}(a) inset), reflecting the stronger hybridization with the Ag(100) conduction bands evident in Fig.~\ref{fig:4_5}(b).

\end{document}